\documentclass[oldversion]{aa}
\usepackage{epsfig}
\usepackage{natbib}
\usepackage{lscape}
\usepackage{wasysym}
\bibpunct{(}{)}{;}{a}{}{,}
\begin{document}

\title{Do stellar magnetic cycles influence the measurement of precise radial velocities?\thanks{Based on 
observations collected at the La Silla Parana Observatory,
ESO (Chile) with the HARPS spectrograph at the 3.6-m telescope (ESO runs ID 
072.C-0096      
073.D-0038    
074.D-0131       
075.D-0194       
076.D-0130      
078.D-0071      
079.D-0075      
080.D-0086
081.D-0065).  Tables 5 to 8, with the data used for Figs. 2, 3, and 8, are available in electronic form
at the CDS via anonymous ftp to cdsarc.u-strasbg.fr (130.79.128.5)
or via http://cdsweb.u-strasbg.fr/cgi-bin/qcat?J/A+A/ }}

%\subtitle{Stellar metallicity, stellar mass, and giant planets}

\author{
  N.C. Santos\inst{1,2} \and	
  J. Gomes da Silva\inst{1,2} \and
  C. Lovis\inst{3} \and	
  C. Melo\inst{4} 
  }

\institute{
    Centro de Astrof{\'\i}sica, Universidade do Porto, Rua das Estrelas, 
    4150-762 Porto, Portugal
    \and
    Departamento de Matem\'atica Aplicada, Faculdade de Ci\^encias da Universidade do Porto, Portugal
    \and
    Observatoire de Gen\`eve, 51 ch. des Maillettes, 1290 Sauverny, Switzerland
    \and
    European Southern Observatory, Casilla 19001, Santiago 19, Chile
}

%    Physikalisches Institut, University of Bern, Sidlerstrasse 5, 3012 Bern, Switzerland
%    \and

\date{Received XXX; accepted YYY}

\abstract{The ever increasing level of precision achieved by
present and future radial-velocity instruments is opening the way to discovering very
low-mass, long-period planets (e.g. solar-system analogs). 
These systems will be detectable as low-amplitude signals in radial-velocity (RV). 
However, an important obstacle to their detection may be the existence
of stellar magnetic cycles on similar timescales. Here we present the results of a long-term program 
to simultaneously measure radial-velocities and stellar-activity indicators (CaII, H$_{\alpha}$, \ion{He}{i}) 
for a sample of stars with known activity cycles. Our results suggest that all these stellar activity
indexes can be used to trace the stellar magnetic cycle in solar-type stars. Likewise,
we find clear indications that different parameters of the HARPS cross-correlation function (BIS, FWHM, and contrast)
are also sensitive to activity level variations. Finally, we show that, although in a few cases slight correlations or anti-correlations between radial-velocity and the
activity level of the star exist, their origin is still not clear. We can, however, conclude that for our targets {(early-K dwarfs)} we do not
find evidence of any radial-velocity variations induced by variations of the stellar magnetic cycle with amplitudes significantly above $\sim$1\,m/s. 
  }

\authorrunning{Santos et al.}
%\titlerunning{}
\maketitle

\section{Introduction}

Following the discovery of the first extrasolar planet orbiting a solar
type star \citep[][]{Mayor-1995}, a multitude
of other planetary systems have been announced \citep[for a review see][]{Udry-2007}. At first, only very short-period companions were found, something that was quite unexpected
by the theories of giant planet formation.
However, the {impressive} increase in precision of
the current radial-velocity (RV) ``machines'' (the majority of the established 
exoplanets have been discovered by radial-velocity surveys) and the
long baseline of the measurements have already brought to light both
long-period planets, much more similar to the Solar System giants \citep[e.g.][]{Wright-2008},
and very low-mass rocky planets \citep[e.g.][]{Mayor-2009}.

Although extremely efficient, the RV technique is
an indirect method, and it does not allow direct detection of
a planet, but only measures the (gravitational) stellar wobble induced 
by the planetary companion orbiting its sun. As a consequence, one of the problems 
is that one has to assure that the radial-velocity variations
observed are not being caused by some other mechanism unrelated to the presence of a 
low-mass companion.
Phenomena such as stellar pulsation \citep[e.g.][]{Bouchy-2004},
inhomogeneous convection, or spots \citep[][]{Saar-1997,Santos-2000a,Paulson-2002} can prevent us from finding planets (if the perturbation is stronger than the orbital radial-velocity variation);
but perhaps more important, they
might give us false candidates if they produce a periodic signal \citep[][]{Queloz-2001,Bonfils-2007,Huelamo-2008}. 
In other words, the radial-velocity technique is sensitive not
only to the motion of a star around the center of mass of a star/planet
system, but also to possible variations in the structure of the stellar
surface. 

\begin{table*}[t]
\caption[]{Basic data for our stars and observation log.}
\begin{tabular}{lcccccccc}
\hline
%\noalign{\smallskip}
Star     & V & B$-$V & Sp. Type & P$_{rot}$ [days] & First observation & Last observation & \#\,nights & $<$S/N$>$\\
\hline
\object{HD4628 }   	&  5.74  &   0.890 &   K2V     &  39 & 2003-10-26 &   2008-09-05 &   36 &   34\\
\object{HD16160}  	&  5.79  &   0.918 &   K3V     &  48 & 2003-10-26 &   2008-09-05 &   43 &   28\\
\object{HD26965A}  	&  4.43  &   0.820 &   K1V     &  43 & 2003-10-26 &   2008-09-05 &   38 &   43\\
\object{HD32147  }	&  6.22  &   1.049 &   K3V     &  47 & 2003-10-26 &   2008-09-05 &   35 &   24\\
\object{HD152391}	&  6.65  &   0.749 &   G8V     &  11 & 2003-06-18 &   2008-09-11 &   38 &   36\\
\object{HD160346}	&  6.53  &   0.959 &   K3V     &  37 & 2004-02-10 &   2008-09-11 &   31 &   29\\
\object{HD216385}	&  5.16  &   0.487 &   F7IV     & 7 &  2003-10-26 &   2008-09-11 &   29 &   56\\
\object{HD219834A}	&  5.20  &   0.787 &   G6/G8IV & 42  & 2003-10-26 &   2008-07-09 &   25 &   48\\
\hline
\end{tabular}
\label{tab:sample}
\end{table*}

This matter is getting more important every day. 
One clear output of the increase in precision of the current
radial-velocity surveys is the ability to find low-mass and long-period planets,
that are able to induce long-period but low-amplitude radial-velocity signals.
These expected Jupiter-analogs might be easily found e.g. with
instruments like HARPS \citep[][]{Pepe-2002} and with future instrumentation
like the projected ESO instruments ESPRESSO\footnote{http://espresso.astro.up.pt}
and CODEX \citep[][]{Pasquini-2008}. 
But if in these cases stellar rotational effects are not potentially 
dangerous (rotational periods are much shorter than the
long-period Jupiter orbit), another almost quantitatively unknown
effect might threaten these detections: stellar activity cycles.

It is now well known that many solar type stars have important
magnetic activity cycles \citep[][]{Baliunas-1995}, similar
to the solar 22-year period (11+11 period spot cycle). 
These cycles have timescales comparable to the orbital periods of
real Jupiter like planets. In this sense, RV measurements of 
the Sun during a 5-year period have revealed that ``our'' star might be 
stable up to a precision of a few m/s \citep[][]{McMillan-1993}. 
However, no tests have been done regarding other solar-type stars.
Some effect of the activity level on the radial-velocity might indeed be expected.
It is known that magnetic fields are able to change the convection patterns 
(e.g. inhibiting convection), thus changing line bisectors
and line shifts \citep[][]{Dravins-1982}.
But with the exception of a very few cases \citep[][for Barnard's star, an M dwarf]{Kurster-2003},
it is not really known or accurately explored,
to which level the line asymmetries induced by changes in
the convection patterns througout the stellar magnetic cycles can 
influence the measurement of precise radial velocities.

In this paper we present a study of the relation between long-term chromospheric activity variations, 
RV, and different line profile indicators. In Sect.\,\ref{sec:sample} we present our sample and observations, and
in Sect.\,\ref{sec:parameters} we derive precise atmospheric parameters and masses for our stars.  
In Sect.\,\ref{sec:indexes} we then present the methodology used to derive the stellar activity level of
our stars from each obtained spectrum. The analysis of the data and the presentation of the results is done in
Sect.\,\ref{sec:results}. We conclude in Sect.\,\ref{sec:conclusions}.

\section{Sample and observations}
\label{sec:sample}

The choice of our targets was based on the sample of stars followed within the Mount Wilson project to study the magnetic cycle 
variations in nearby FGK stars \citep[][]{Vaughan-1978,Baliunas-1995}. In particular, we based our choice on the sample 
presented in \citet[][]{Baliunas-1995}. From the stars and results presented by these authors, we first took those showing clear 
activity cycles. Then, only objects at a declination south of $+$10 degrees were considered, in order to be able to follow them 
from the La Silla observatory (the site of HARPS) for several months each year. One star that presented a particularly stable 
activity level was also chosen as the standard (\object{HD\,216385}). 

In Table\,\ref{tab:sample} we list the stars in our sample with their V magnitudes, B-V colors, and spectral types. 
The values were taken from the Hipparcos catalog \citep[][]{ESA-1997}. Rotational periods were taken from \citet[][]{Baliunas-1996}. As we can see from the table, most stars are 
late-G or early-K dwarfs, with the exception of \object{HD\,216385} (late F) and \object{HD\,219834A} (likely a subgiant) -- see also Table\,\ref{tab:parameters}. 

To allow for a continuous follow-up of each target, observations were done in service mode starting from September 2003 (ESO period 72), 
just when HARPS became available. All stars were followed until September 2008 (ESO period 81). No more follow-up was then possible 
due to the end of service-mode observations at La Silla. For \object{HD\,152391}, one extra measurement done during the commissioning of HARPS (June 2003) was added.

In general, each star was observed in 5 different epochs during each observing season (6 months). An effort was made to spread the observations in time as much as 
possible. This was important to allow us to average any strong variations due to rotational modulation effects out. We are 
interested in studying the effects of long term magnetic cycle variations.

Every measurement was done with the high-precision simultaneous calibration mode. Though the use of the Thorium-Argon simultaneous 
calibration may complicate the subtraction of scattered light over the whole image (specially in the blue part of the spectrum where the 
spectral orders are separated by only a few pixels), this mode was judged important since we need m/s precision in our measurements. 

\begin{table*}[t!]
\caption[]{Stellar parameters derived for the stars in our sample.}
\begin{tabular}{lcccrccccc}
\hline
%\noalign{\smallskip}
Star     & T$_{\mathrm{eff}}$ & $\log{g}_{spec}$        & $\xi_{\mathrm{t}}$ & \multicolumn{1}{c}{[Fe/H]} & N(\ion{Fe}{i}/\ion{Fe}{ii}) & $\sigma(\ion{Fe}{i}/\ion{Fe}{ii}$) & $\log{g}_{hipp}$ & Mass       \\
         & [K]                &  [cm\,s$^{-2}$]  &  [km\,s$^{-1}$]    &        &             &                  & [cm\,s$^{-2}$]  & [M$_\odot$] \\
\hline
\object{HD4628 }   	& 5044$\pm$52	 & 4.54$\pm$0.36     & 0.53$\pm$0.18	& $-$0.32$\pm$0.11 & 257/31  &  0.11/0.16 &4.61 &0.72 \\
\object{HD16160}  	& 4921$\pm$103	 & 4.54$\pm$0.54     & 0.46$\pm$0.43	& $-$0.16$\pm$0.19 & 258/32  &  0.18/0.24 &4.58 &0.69 \\
\object{HD26965A}  	& 5136$\pm$50	 & 4.46$\pm$0.30     & 0.44$\pm$0.15	& $-$0.32$\pm$0.09 & 258/32  &  0.09/0.13 &4.42 &0.65 \\
\object{HD32147  }	& 4902$\pm$136	 & 4.33$\pm$0.73     & 0.60$\pm$0.54	& 0.21$\pm$0.21    & 259/34  &  0.19/0.34 &4.62 &0.79\\
\object{HD152391}	& 5474$\pm$25	 & 4.46$\pm$0.19     & 1.01$\pm$0.04	& $-$0.04$\pm$0.06 & 256/33  &  0.06/0.08 &4.56 &0.92 \\
\object{HD160346}	& 4998$\pm$68	 & 4.44$\pm$0.38     & 0.81$\pm$0.17	& $-$0.08$\pm$0.14 & 258/32  &  0.14/0.18 &4.63 &0.78\\
\object{HD216385}	& 6371$\pm$30	 & 4.24$\pm$0.34     & 1.63$\pm$0.04	& $-$0.13$\pm$0.06 & 230/36  &  0.06/0.12 &4.06 &1.36\\
\object{HD219834A}	& 5530$\pm$41	 & 3.95$\pm$0.17     & 1.06$\pm$0.04	&  0.13$\pm$0.11   & 257/34  &  0.11/0.07 &3.96 &1.25 \\
\hline
\end{tabular}
\label{tab:parameters}
\end{table*}

Stellar oscillations can induce significant radial-velocity signals on timescales of a few minutes. In order to average the stellar oscillation 
modes out \citep[e.g.][]{Santos-2004a}, each measurement was done with a total exposure time of 15 minutes. 
For the brightest stars in our sample, this implies several shorter exposures to avoid CCD saturation.

\begin{figure}[b!]
\resizebox{\hsize}{!}{\includegraphics{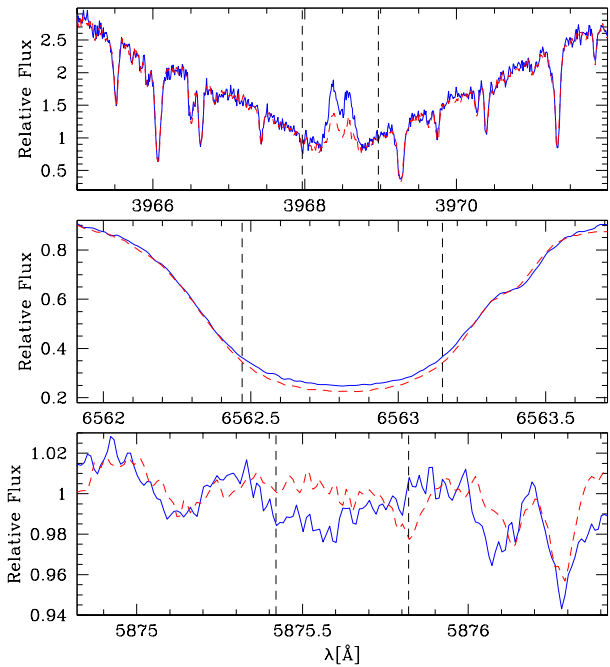}}
\caption{Comparison of two spectra of HD4628 near its maximum (blue continuous line) and minimum (red dashed line) activity level in the regions of
the \ion{Ca}{ii} H line (top), the H$_\alpha$ line (middle), and the \ion{He}{i} D3 line (lower panel). The vertical dashed lines denote the regions
used to derive the activity indexes.} 
\label{fig:lines}
\end{figure}

Individual RVs were derived with the HARPS pipeline. The velocities were derived using the cross-correlation function (CCF)
method. The bisector inverse slope (BIS) was derived using the methodology described in \citet[][]{Queloz-2000}. Other parameters
of the CCF, such as the contrast and FWHM, were also computed.

\section{Spectroscopic stellar parameters and masses}
\label{sec:parameters}

With the combined HARPS spectra, we derived stellar parameters and metallicities for the sample stars 
using the methodology and line-lists described in \citet[][]{Santos-2004b} and \citet[][]{Sousa-2008}, respectively. Equivalent 
widths of individual \ion{Fe}{i} and \ion{Fe}{ii} lines were measured with the automatic code ARES\footnote{http://www.astro.up.pt/$\sim$sousasag/ares}. 
A grid of \citet[][]{Kurucz-1993} model atmospheres was adopted together with the 2002 version of the radiative transfer code 
MOOG \citep[][]{Sneden-1973}. We point the readers to \citet[][]{Santos-2004b} for more details on the technique. In Table\,\ref{tab:parameters} 
we list the final derived values, together with the number of \ion{Fe}{i} and \ion{Fe}{ii} lines used and their dispersion.
For comparison, the astrometric surface gravity was also derived using Eq.\,1 in \citet[][]{Santos-2004b}. Both estimates of the surface
gravity agree well within the errors. 

Stellar masses were also derived by interpolating the theoretical isochrones of \citet{Schaller-1992}, 
\citet[][]{Schaerer-1993a}, and \citet{Schaerer-1993b}, using $M_{V}$ computed using Hipparcos 
parallaxes \citep[][]{ESA-1997}, a bolometric correction from \citet[][]{Flower-1996}, and the
T$_{\mathrm{eff}}$ obtained from the spectroscopy. The results are also listed in Table\,\ref{tab:parameters}.
We estimate that the uncertainties are around 10\% because of errors in the input parameters and also because of different systematic effects \citep[][]{Fernandes-2004}.

A look at Table\,\ref{tab:parameters} shows that all but one of the targets are main sequence dwarfs. The exception 
is HD\,219834\,A, most likely a subgiant.

\section{Activity indexes}
\label{sec:indexes}

From each individual spectrum, stellar activity indexes were derived using three different indicators: the \ion{Ca}{ii} H and K lines, the
H$_\alpha$ line, and the \ion{He}{i} D3 line at 5875.6\,\AA. All activity indexes were derived following the general procedures described 
in \citet[][for the \ion{Ca}{ii} lines]{Santos-2000a} and \citet[][for H$_\alpha$ and \ion{He}{i}]{Boisse-2009}. 
All the activity level derivations used the pipeline reduced and wavelength calibrated 2-dimensional spectra. 
Individual errors were derived with the number of counts in each wavelength region.

\begin{table*}[t]
\caption[]{Average chromospheric activity indexes derived using the \ion{Ca}{ii} H and K lines, with the values presented in \citet[][]{Baliunas-1995} presented for comparison.}
\begin{tabular}{lccc}
\hline
%\noalign{\smallskip}
Star     & $<S_{MW}>^{(ours)}$  & $<S_{MW}>^{(Baliunas)}$ & $<\log{R'_{HK}}>^{(ours)}$\\
\hline
\object{HD4628 }       &  0.22  & 0.230  &  $-$4.89     \\
\object{HD16160}      &  0.23  & 0.226  &  $-$4.88    \\
\object{HD26965A}      &  0.19  & 0.206  &  $-$4.92     \\
\object{HD32147  }    &  0.28  & 0.286  &  $-$4.95   \\
\object{HD152391}    &  0.39  & 0.393  &  $-$4.44     \\
\object{HD160346}    &  0.31  & 0.300  &  $-$4.78    \\
\object{HD216385}    &  0.15  & 0.142  &  $-$5.00    \\
\object{HD219834A}    &  0.16  & 0.155  &  $-$5.03  \\
\hline
\end{tabular}
\label{tab:sindex}
\end{table*}

\begin{figure*}[t!]
\resizebox{\hsize}{!}{\includegraphics[bb = 21 505 586 643]{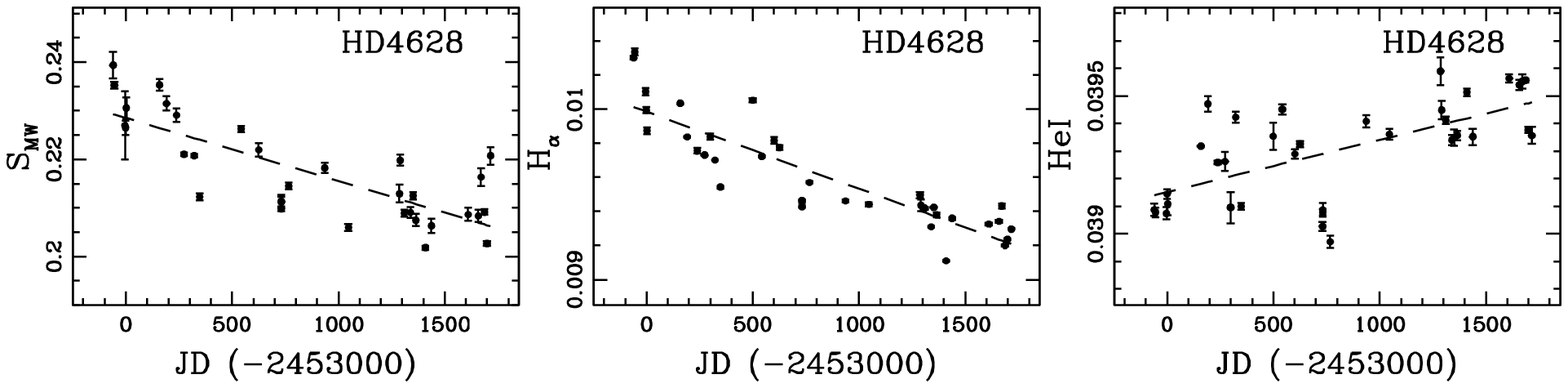}}
\resizebox{\hsize}{!}{\includegraphics[bb = 21 505 586 643]{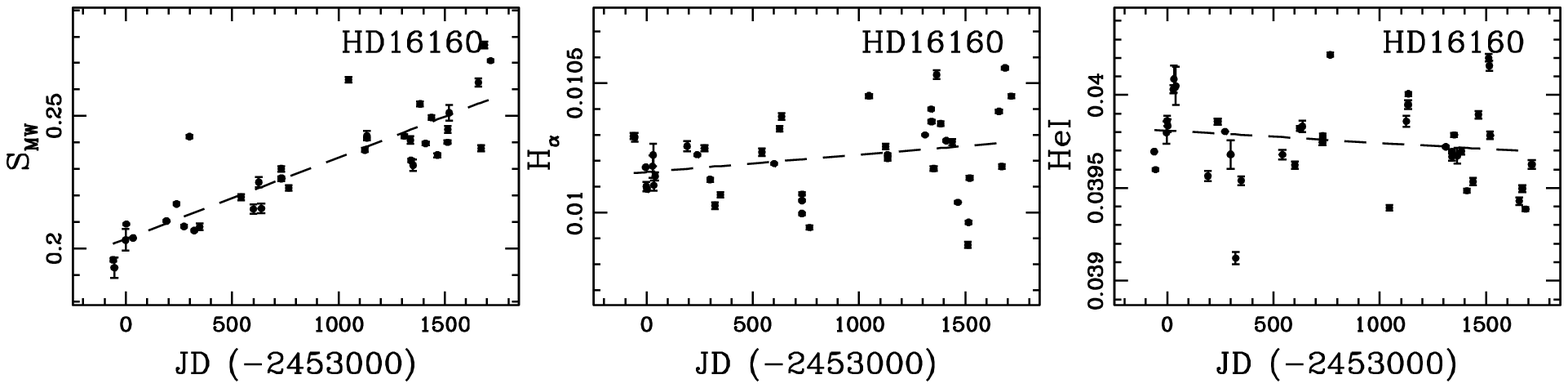}}
\resizebox{\hsize}{!}{\includegraphics[bb = 21 505 586 643]{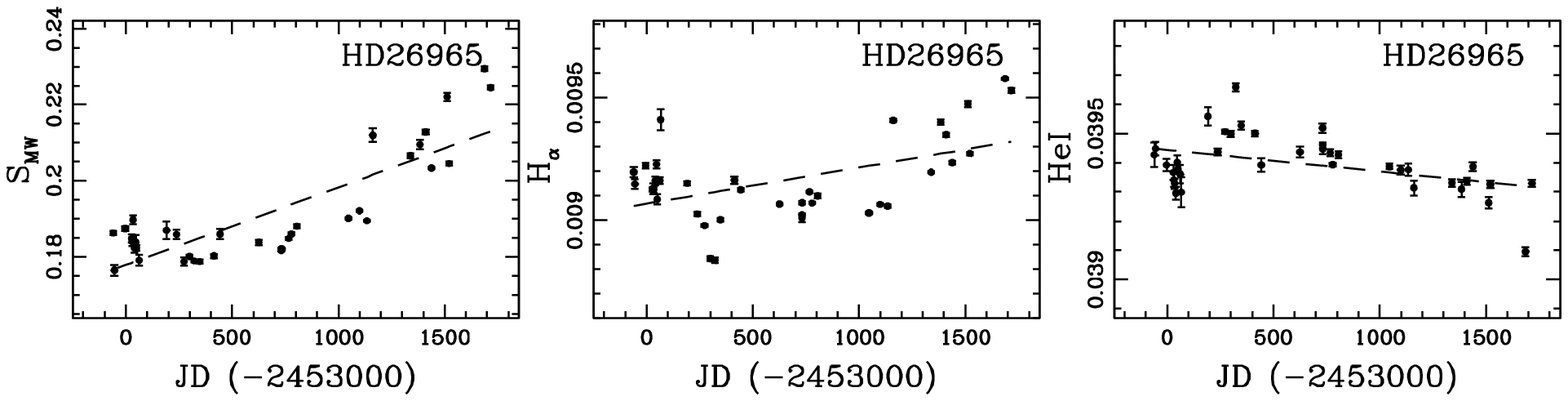}}
\resizebox{\hsize}{!}{\includegraphics[bb = 21 505 586 643]{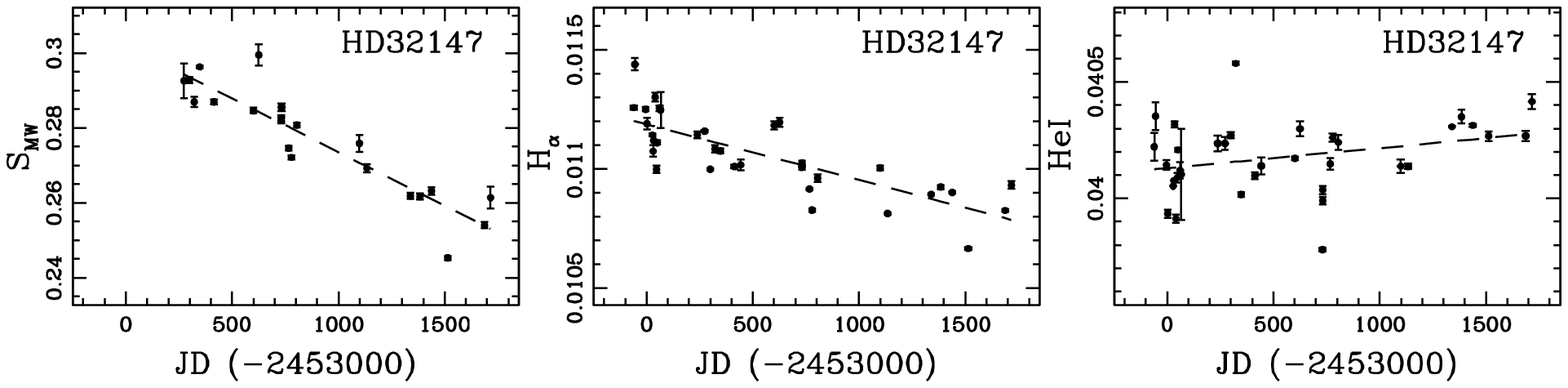}}
\caption{Activity indexes for HD\,4628, HD\,16160\, HD26965A, and HD\,32146 as a function of time. Left panels denote 
variations in the Mount Wilson H and K  ``S'' index, middle panels the H$_\alpha$ index, and the right panels the \ion{He}{i} index. 
The dashed line represents a linear fit to the data. For comparison reasons, the x-scale was set constant in all the plots.} 
\label{fig:activity1}
\end{figure*}

\begin{figure*}[t!]
\resizebox{\hsize}{!}{\includegraphics[bb = 21 505 586 643]{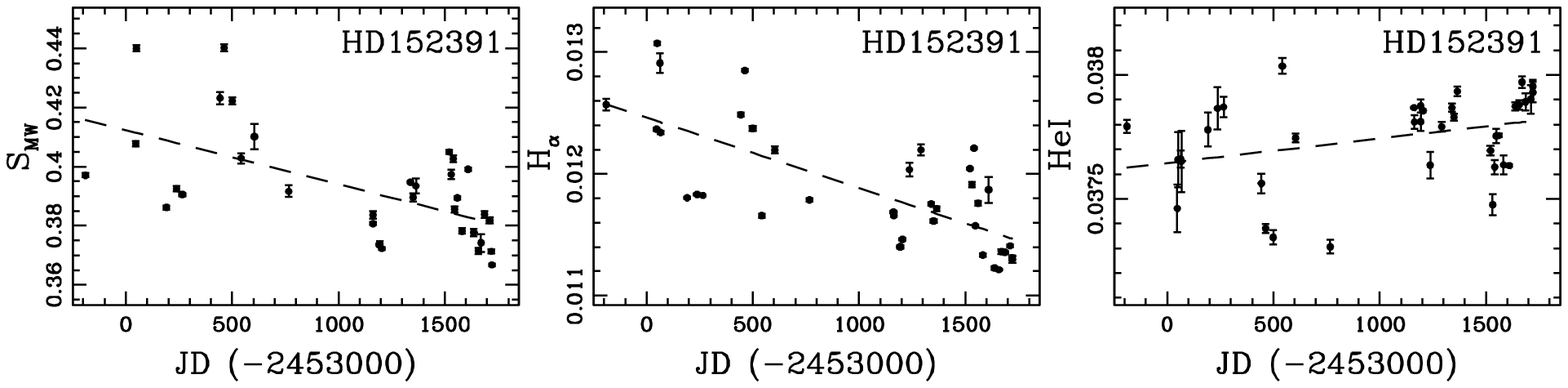}}
\resizebox{\hsize}{!}{\includegraphics[bb = 21 505 586 643]{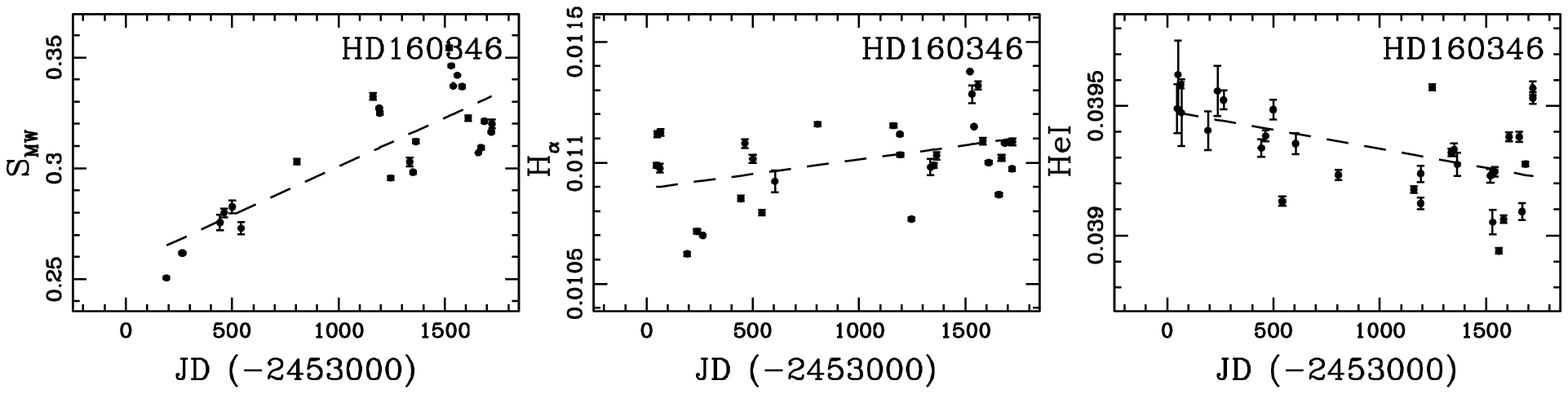}}
\resizebox{\hsize}{!}{\includegraphics[bb = 21 505 586 643]{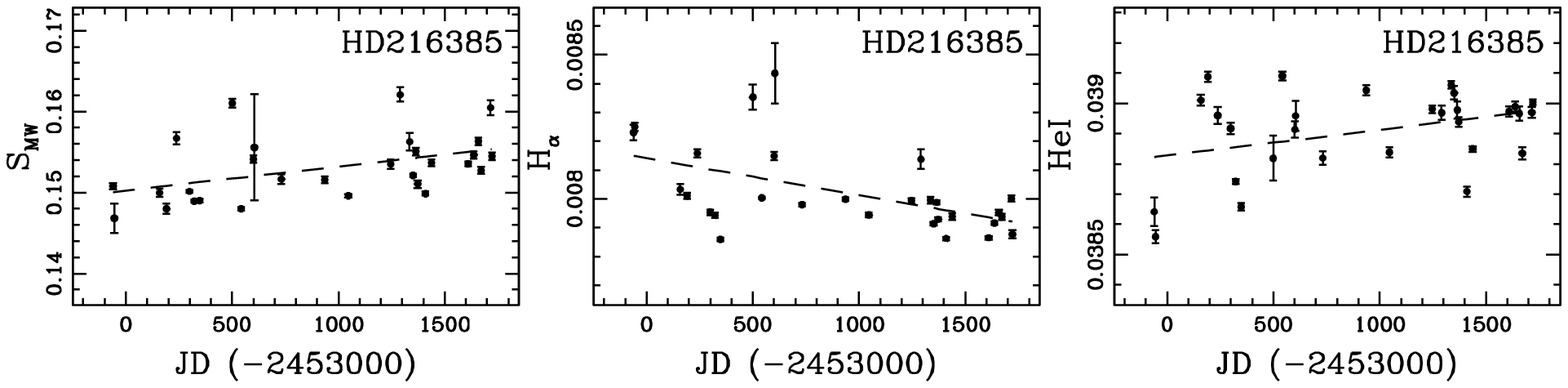}}
\resizebox{\hsize}{!}{\includegraphics[bb = 21 505 586 643]{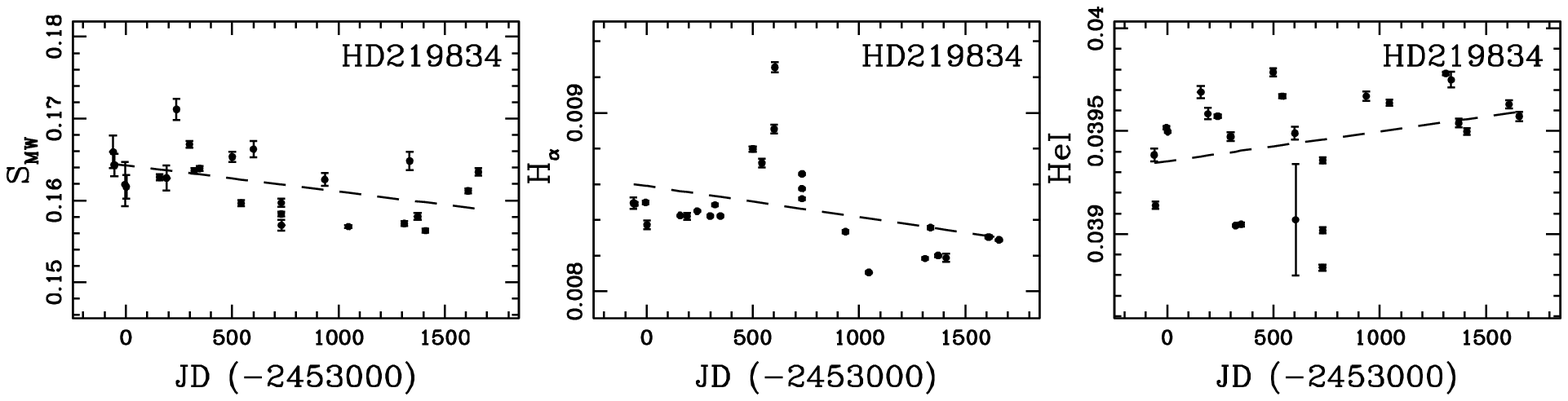}}
\caption{Same as Fig.\,\ref{fig:activity1} for HD\,152391, HD\,160346, HD\,216385, and HD\,219834A.} 
\label{fig:activity2}
\end{figure*}

The \ion{Ca}{ii} H and K ``S$_{HARPS}$" index was computed by dividing the sum of the flux in two 1\,\AA\,wide intervals in the center of each
line (at 3933.66 and 3968.47\,\AA, respectively), by the flux in two 20\,\AA\, wide reference windows centered on 3900.0 and 4000.0\,\AA. A 
weighted sum took the photon noise errors derived as $\sqrt{N}$ into account  (with N the number of counts). Once all the spectra 
were analyzed, 
an average S$_{HARPS}$ value was derived for each star. Its value was then used to calibrate the S$_{HARPS}$ values to the 
Mount-Wilson scale. For this calibration we used the S$_{MW}$ values listed in \citet[][]{Baliunas-1995}\footnote{The 
calibration yealds S$_{MW}$ = 30.985$\times$S$_{HARPS}$ + 0.042}  -- see Table\,\ref{tab:sindex}. Finally, using the relation in \citet[][]{Noyes-1984},
we could derive the values of the \ion{Ca}{ii} flux corrected for the photospheric flux, $\log{R'_{HK}}$. 

In Table\,\ref{tab:sindex} we list the average values for each of our targets (both for S$_{MW}$ and $\log{R'_{HK}}$), together with 
the values for the S$_{MW}$ listed
in \citet[][]{Baliunas-1995}. As can be seen from the table, the values derived here agree perfectly with those
derived by Baliunas et al. This also reflects that, during the 5 years of measurements, we could follow a significant
part of the stellar magnetic cycle.
Except for HD\,152391, all the stars seem to be in the low-activity side of the Vaughan-Preston gap \citep[][]{Vaughan-1980},
and are thus similar to the Sun regarding their activity level.

The H$_\alpha$ index was derived by dividing the flux in the central 0.678\,\AA\, of the H$_\alpha$ line (6562.808\,\AA) by
the flux in two reference windows between 6545.495-6556.245 and 6575.934-6584.684\,\AA. Finally, the
\ion{He}{I} D3 index was derived by dividing the flux in the 0.4\,\AA\, central region of the 5875.62\,\AA\, line by
the flux in two 5\,\AA\, wide reference windows in each side of the line, centered on 5869 and 5881\,\AA.

While the  \ion{Ca}{ii} H and K lines and H$_\alpha$ line typically present a re-emission at the center associated 
with activity phenomena, the depth of the \ion{He}{i} D3 line increases in solar plages \citep[][]{Landman-1981}. A correlation is thus expected 
to be present between the \ion{Ca}{ii} H and K index and the H$_\alpha$ index, while these two indexes will in principle be anti-correlated 
with the \ion{He}{i} D3 activity indicator. This can be seen in Fig.\,\ref{fig:lines}, though from this plot is is also clear that the two
former lines are more sensitive to activity variations than the latter one.
This is partially caused by the weakness of the \ion{He}{i} D3 line, which it is dilluted by several atomic lines in the same region, and
affected by telluric absorption features \citep[][]{Danks-1985,Saar-1997b}.

The wavelength region where the \ion{Ca}{ii} H and K lines are present (blue) often has a low S/N, making it difficult in some cases to 
derive a reliable activity index. Beyond this, and because the spectra were collected with a simultaneous
ThAr wavelength calibration spectrum, contamination from the nearby ThAr order may in some cases alter the results. To avoid this, only spectra with an S/N$>$20 near 4000\,\AA\ were used to derive the  \ion{Ca}{ii} H and K ``S$_{HARPS}$" index.

\section{Results}
\label{sec:results}

\subsection{Activity cycle variations}

In Figs.\,\ref{fig:activity1} and \ref{fig:activity2} we present time series of the three 
activity indexes for the 8 stars in our program\footnote{Data available in online Tables\,5, 6, and 7}. Each point in the figure denotes
the average of the measurements over one observing night (if multiple spectra existed for a given night). The error bars were computed
using the dispersion of the different measurements in case several spectra per
night were taken, or the individual error for nights when one single spectrum was obtained.

A general look at the plots show that several of the stars present clear long-term activity cycle variations, as expected. In general also, 
these variations are seen in all the activity indexes \citep[see also discussions in][]{Livingston-2007,Cincunegui-2007,Meunier-2009}, though the \ion{Ca}{II} S index variations seem clearer and present smaller dispersion. Our data suggest, however, that all these indexes can 
be used to trace long-term activity variations in solar-type stars.

As mentioned above, the \ion{He}{i} index is generally anti-correlated with the H$_{\alpha}$
and \ion{Ca}{ii} H and K indexes (the last two are generally correlated). This is expected from 
studies of the solar active regions \citep[][]{Landman-1981}. The exception for this is the late F dwarf HD\,216385. 
Though not the scope of the present paper, our results thus suggest that in F dwarfs the physics 
of the formation of the \ion{He}{i} line is different from the one seen for {early-K dwarfs}. 
This star presents, however, a very stable activity level througout all the series of measurements, confirming
its stability as shown in \citet[][]{Baliunas-1995} -- it was actually included in our sample as a ``standard''.

Some of the stars show higher frequency structure. This is likely due to the appearance and disappearance of activity 
phenomena such as spots and plages, and their modulation with stellar rotation. In the present paper we are mostly 
interested in long-term variations. We will leave the discussion of these high frequency variations to a different paper.

\object{HD\,219834A} presents a clear spike in the H$_\alpha$ and \ion{He}{i} indexes near JD$\sim$2453600. As shown 
in Sect\,\ref{sec:multiples}, this star is a long-period binary with an eccentricity significantly different from zero. 
A fit to the radial velocities shows that the periastron passage occurs near the above date. We may thus be 
seeing activity induced by the fainter secondary star on HD\,219834A. This issue is beyond the scope of the present paper, 
but will be discussed in a separate work.

\subsection{Multiple stars}
\label{sec:multiples}

Three of the stars in our sample are members of multiple stellar systems. To be able to study the
influence that magnetic cycle variations may have on the measurement of RVs,
we need to subtract the variations caused by the orbital period. In this section we describe
the procedures used in each case. 

\begin{figure}[b!]
\resizebox{\hsize}{!}{\includegraphics{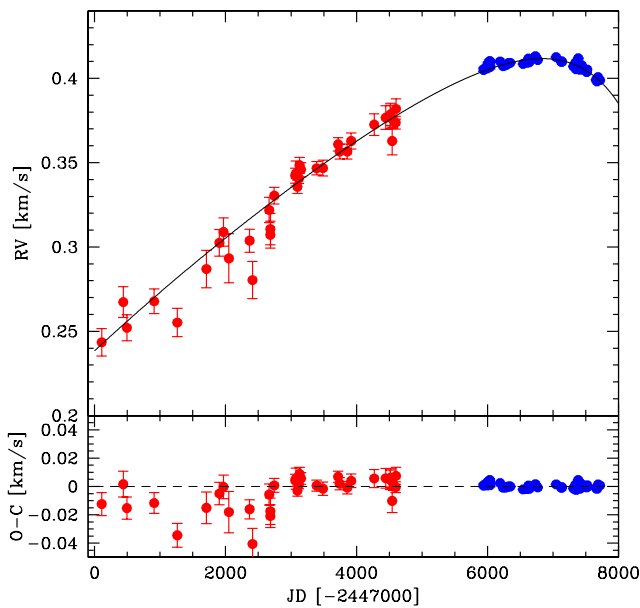}}
\caption{Radial velocities of HD\,16160 and best-fit Keplerian orbit. The lower panel presents the residuals of the fit.} 
\label{fig:16160}
\end{figure}

\begin{table*}[t!]
\caption[]{Orbital elements of the fitted orbits for HD\,16160, HD\,160346, and HD\,219834. See text for more details}
\begin{tabular}{lllll}
\hline
\hline
\noalign{\smallskip}
Parameter & HD\,16160 & HD\,160346 & HD\,219834\,A & Units\\
\hline
$P$		& 21900$\dagger$	& 83.7288$\pm$0.0007		& 2338$\pm$17		&	[d]\\
$T$		& 2\,457\,280$\pm$271	& 2\,454\,590.40$\pm$0.02	& 2\,453\,768$\pm$19	&	[d]\\  %Na ultima juntei um periodo ao T0
$a$		& 14.0			& 0.36				& 4.10			&	[AU]\\
$e$		& 0.75$\dagger$		& 0.2048$\pm$0.0004		& 0.184$\pm$0.005	&	\\
$V_r$ (HARPS)	& 25.53$\pm$0.05	& 21.941$\pm$0.002		& 10.56$\pm$0.03	&	[km\,s$^{-1}$]\\
$\omega$	& 144$\pm$2		& 140.1$\pm$0.1			& 213.9$\pm$0.7		&	[degr] \\ 
$K_1$		& 1\,052$\pm$194	& 5\,691$\pm$3			& 6017$\pm$11		&	[m\,s$^{-1}$] \\
$\sigma(O-C)$	& 1.51			& 7.11				& 25.8			&	[m\,s$^{-1}$]  \\    
$N$		& 78			& 33				& 25			&	\\
$m_2\,\sin{i}$	& 74			& 101				& 448			&	[M$_{\mathrm{Jup}}$]\\
\noalign{\smallskip}
\hline
\end{tabular}
\newline
$\dagger$ Fixed according to \citet[][]{Golimowski-2000}
\label{tab:orbits}
\end{table*}

\begin{figure}[b!]
\resizebox{\hsize}{!}{\includegraphics{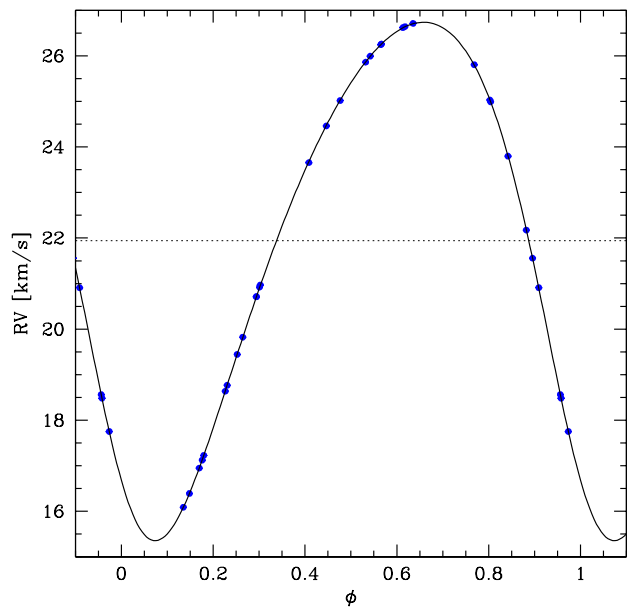}}
\caption{Phase folded radial-velocities of HD\,160346 and best-fit Keplerian orbit.} 
\label{fig:160346}
\end{figure}

\begin{figure}[t!]
\resizebox{\hsize}{!}{\includegraphics{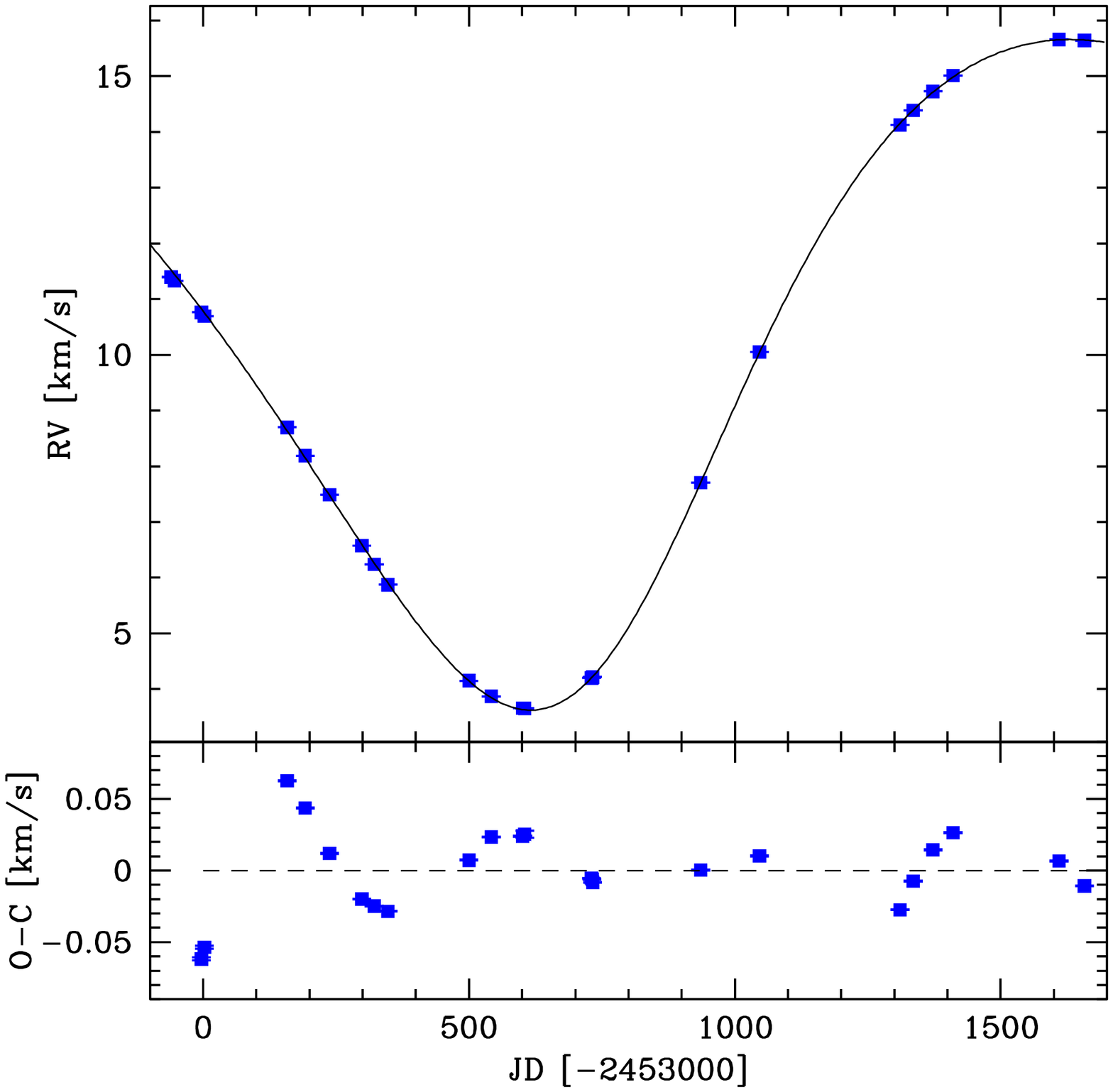}}
\caption{Radial velocities of HD\,219834A and best-fit Keplerian orbit. The lower panel presents the residuals of the fit.} 
\label{fig:219834}
\end{figure}

\subsubsection{HD\,16160}

\object{HD\,16160} (Gl\,105\,A) is member of a known triple system, with Gl\,105\,C orbiting Gl\,105\,A with
a period of $\sim$60\,years, orbital eccentricity of $\sim$0.75, and with semimajor axis of $\sim$15\,AU \citep[][]{Golimowski-2000}.
Since our data only cover a small part of the orbital phase, we used the radial-velocity measurements of this star 
presented by Golimowski et al.\footnote{Taken directly from their postscript figure.} ,
together with our own measurements, to fit a Keplerian function to the system and obtain a global orbital solution (see Table\,\ref{tab:orbits} and Fig.\,\ref{fig:16160}). 
In this process we fixed the orbital period and eccentricity to the values mentioned above.
We used the residual radial velocities of this fit for the rest of the paper.

Leaving all the parameters free we find a slightly better solution with an orbital period $\sim$11\,000 days, but this solution is not significantly 
better than the one adopted. Since our data, together with the one of \citet[][]{Golimowski-2000}, do not cover one entire orbital period, 
we have no way to constrain the orbital solution using only radial velocities.

\subsubsection{HD\,160346}

\citet[][]{Halbwachs-2003} used CORAVEL radial-velocity measurements with Hipparcos \citep[][]{ESA-1997} astrometry to
conclude that \object{HD\,160346} (GJ\,688) is a spectroscopic binary (SB1) with a period of 83.7\,days, eccentricity of 0.20, and
mass ratio $M_2/M_1$ of 0.5. A Keplerian fit to the HARPS radial velocities perfectly confirm this result (Table\,\ref{tab:orbits} and Fig.\,\ref{fig:160346}). The residuals around this Keplerian fit will be used for the rest of this work.

\subsubsection{HD\,219834A}

\object{HD\,219834A} is a known hierarchical triple system, where the A component has been identified as a spectroscopic binary SB1 \citep[][]{Duquennoy-1991,Tokovinin-1997}. Our HARPS data perfectly confirm (and refine) the previously determined orbital parameters of the system. Our radial-velocities
are best fit with a Keplerian function with a period of 2337\,days, eccentricity 0.18, and semi-amplitude K=6.018\,km\,s$^{-1}$ (Table\,\ref{tab:orbits} and Fig.\,\ref{fig:219834}). This signal is compatible with a low mass (0.43\,M$_\odot$ -- minimum mass) stellar companion orbiting the slightly 
evolved 1.25\,M$_\odot$ star HD\,219834\,A.

Unfortunately, the high residual velocities around the orbital fit (rms of $\sim$25\,m\,s$^{-1}$), together with their structure,
strongly suggest that we are observing a blended spectrum and that HD\,219834\,A is an SB2. Alternatively, additional companions could be present in the system. Better fits are indeed obtained with 2 or 3 Keplerian functions, but because our data barely covers one orbital period it does not allow us to find any convincing result. 

These facts make it impossible to use the measurements of this star to study 
long-term and low-amplitude RV variations. We have thus excluded the data of HD\,219834\,A for the rest of the paper.

\subsection{HD\,152391}

\begin{figure}[t!]
\resizebox{\hsize}{!}{\includegraphics{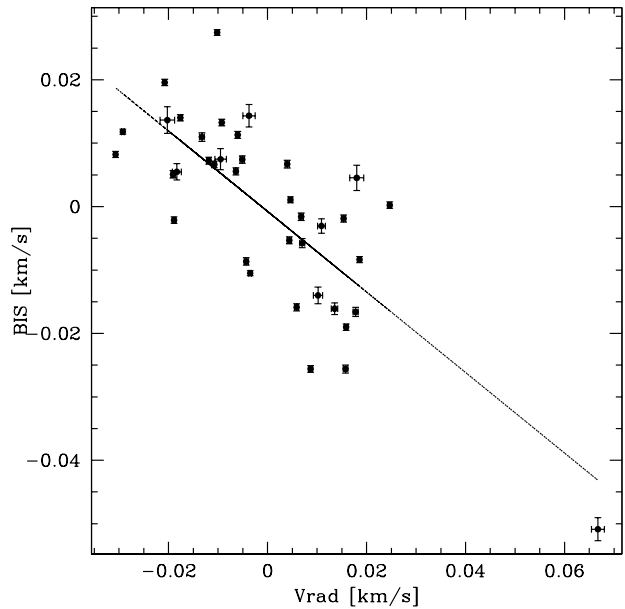}}
\caption{Radial velocities vs. BIS for HD\,152391.} 
\label{fig:bis}
\end{figure}

HD\,152391 is the most active star in our sample. As expected, a look at the obtained RVs reveal
a high noise level close to 18\,m\,s$^{-1}$. A clear correlation between the velocities and BIS is also
seen (Fig.\,\ref{fig:bis}). As for other similar stars \citep[such as HD\,166435 --][]{Queloz-2000}, this seems
to indicate that the observed radial-velocity variations are being induced by photospheric
features like spots. In contrast to the case of \object{HD\,166435}, however, we could not find any clear periodicity
in the data.

We tried to correct the radial-velocities using the relation between BIS and RV, a procedure already
successfully used by \citet[][]{Melo-2007}. However,  the rms around the average velocity only
decreased slightly to $\sim$12.5\,m\,s$^{-1}$. This high value makes the search for any 
low-amplitude and long-term signal very difficult. We thus decided to keep this star out of
the rest of the discussion.

\subsection{Activity, radial-velocity, and CCF parameters}

After removing HD\,152391 and HD\,219834A from the list (see discussion above), we are left
with 6 objects (HD\,4628, HD\,16160, HD\,26965A, HD\,32147, HD\,160346, and HD\,216385) for which we can study the influence 
of activity cycle variations on the measured radial velocities. 

We searched the literature for references to multiplicity among these targets. 
Except for HD\,16160 and HD\,160346 (see discussion above), none of the remaining stars has any reference for a short-period (up to a few years) companion. HD 26965A is member of a known triple system, with 
a long period of $\sim$8\,000\,years \citep[][]{Heintz-1974}. HD\,216385 is also known to be member of a common proper motion 
pair with the second star situated 250 arcsec away \citep[][]{Lepine-2007}; at a distance of $\sim$26\,pc, this corresponds
to a projected separation of $\sim$6500\,AU. For HD\,4628, HD\,26965A, HD\,32147, and HD\,216385 we thus have no
indication that radial-velocity variations induced by a stellar companion can be detectable in our 5-year baseline.

\begin{figure*}[t!]
\center
\resizebox{7.7truecm}{!}{\includegraphics[bb = 25 170 586 707]{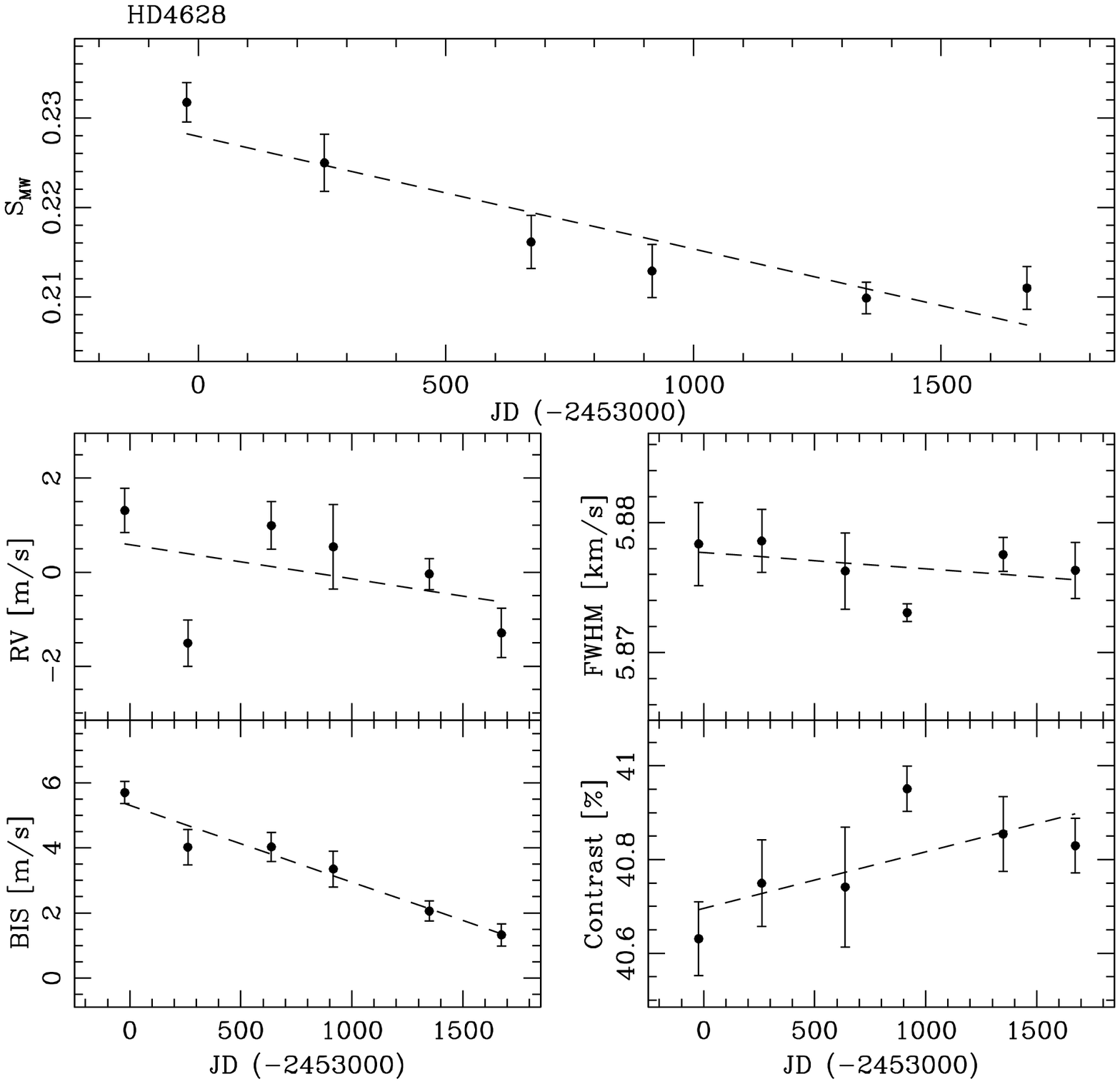}}
\resizebox{7.7truecm}{!}{\includegraphics[bb = 25 170 586 707]{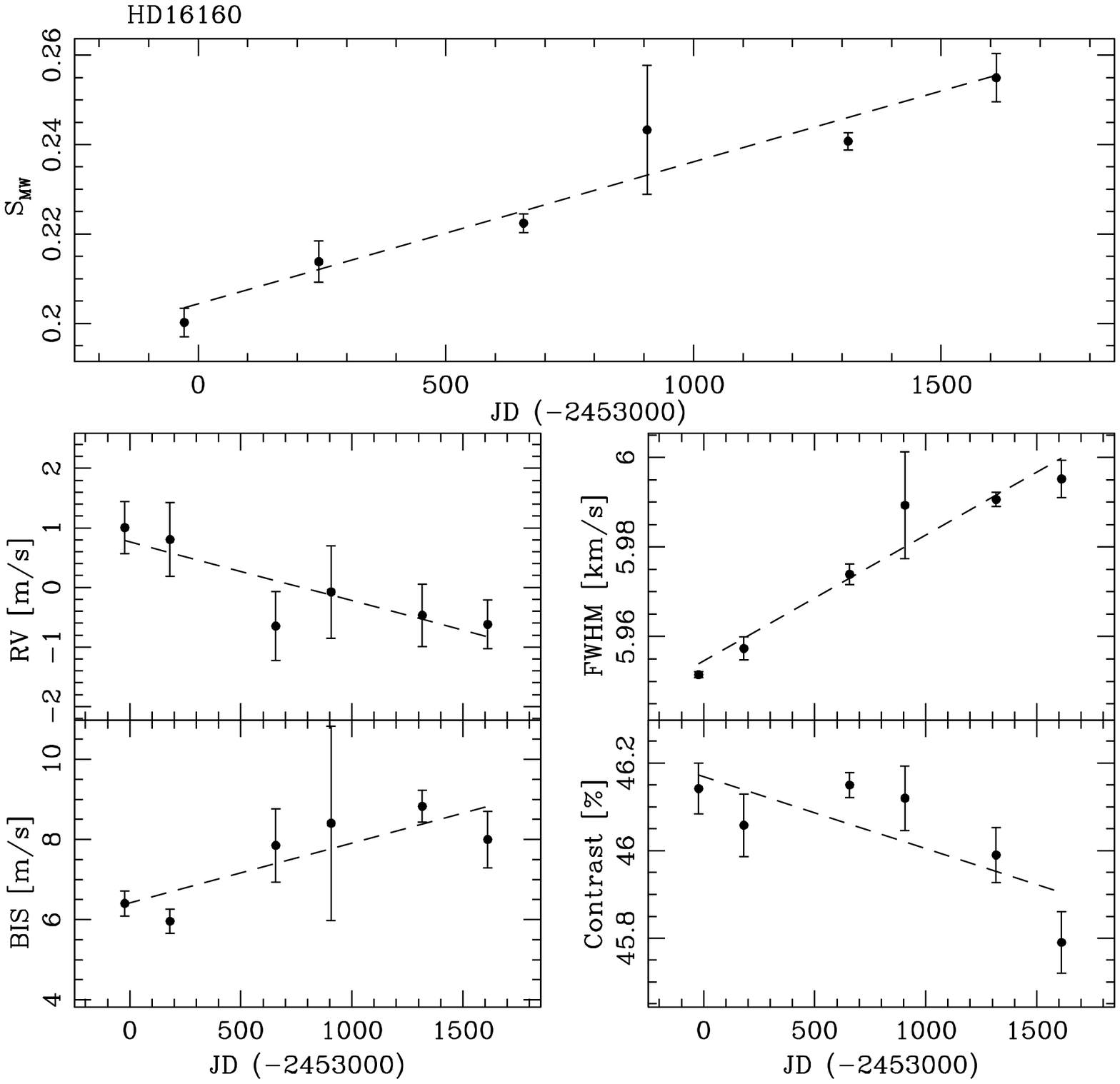}}\\
\resizebox{7.7truecm}{!}{\includegraphics[bb = 25 170 586 707]{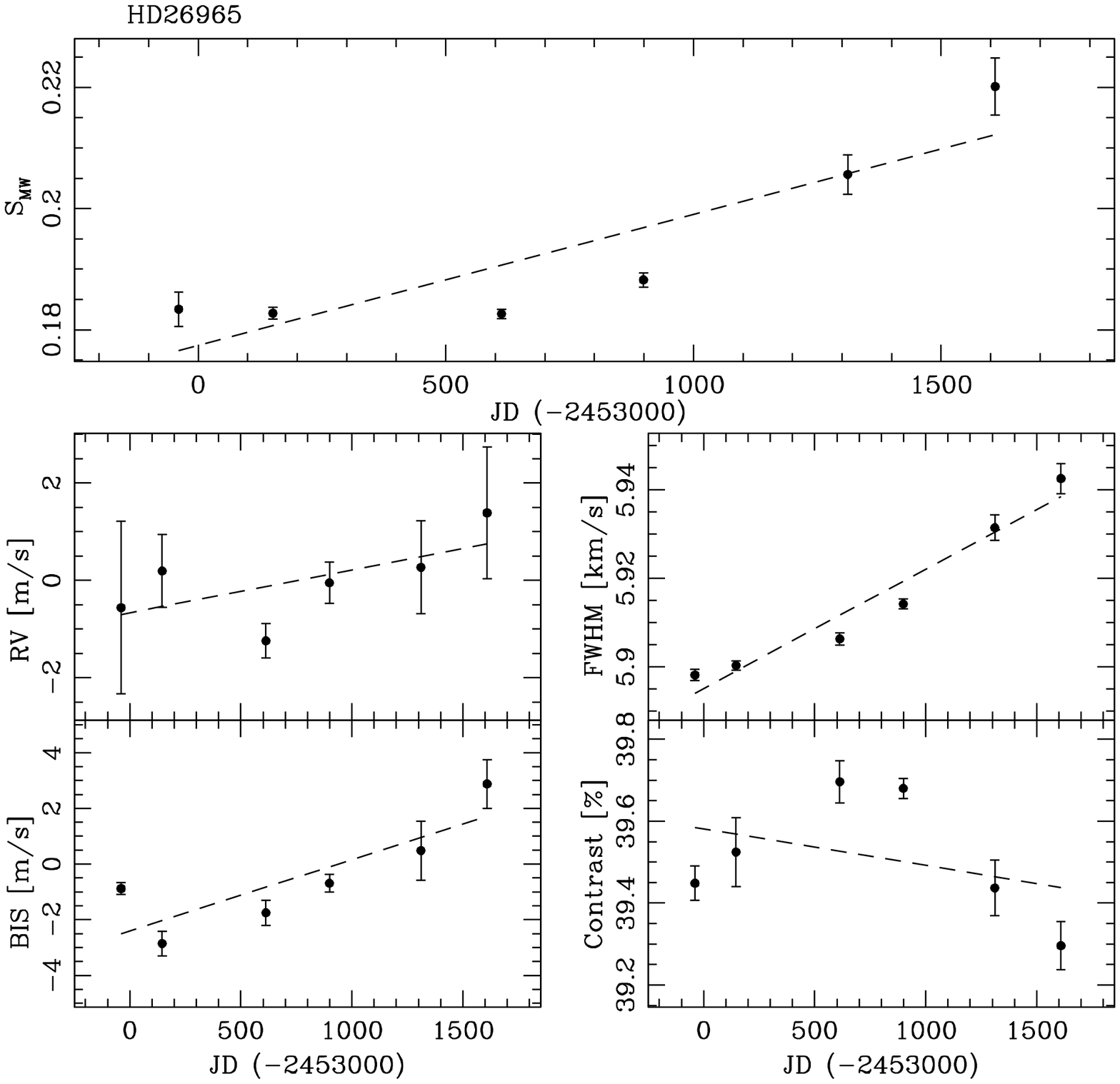}}
\resizebox{7.7truecm}{!}{\includegraphics[bb = 25 170 586 707]{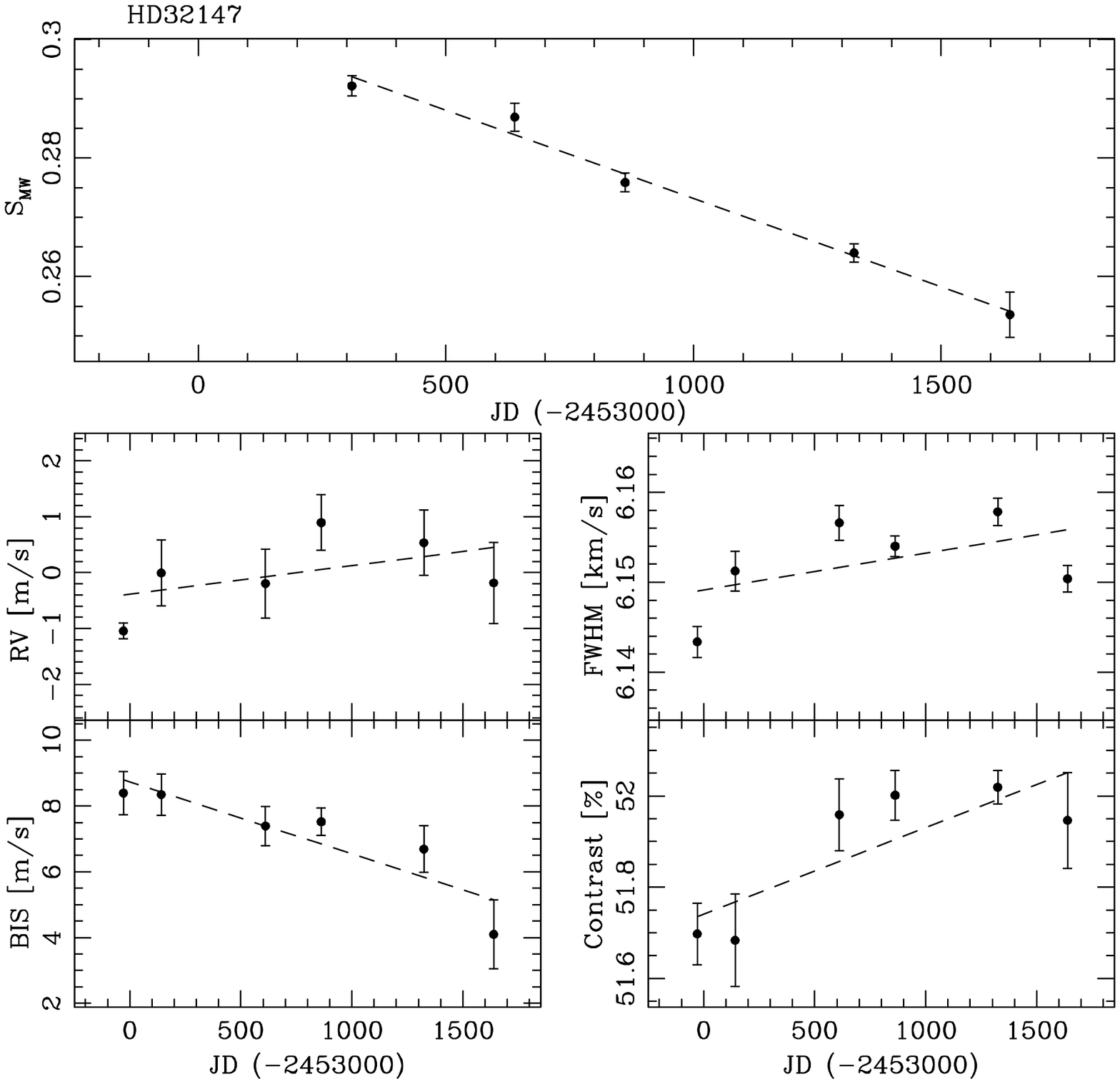}}\\
\resizebox{7.7truecm}{!}{\includegraphics[bb = 25 170 586 707]{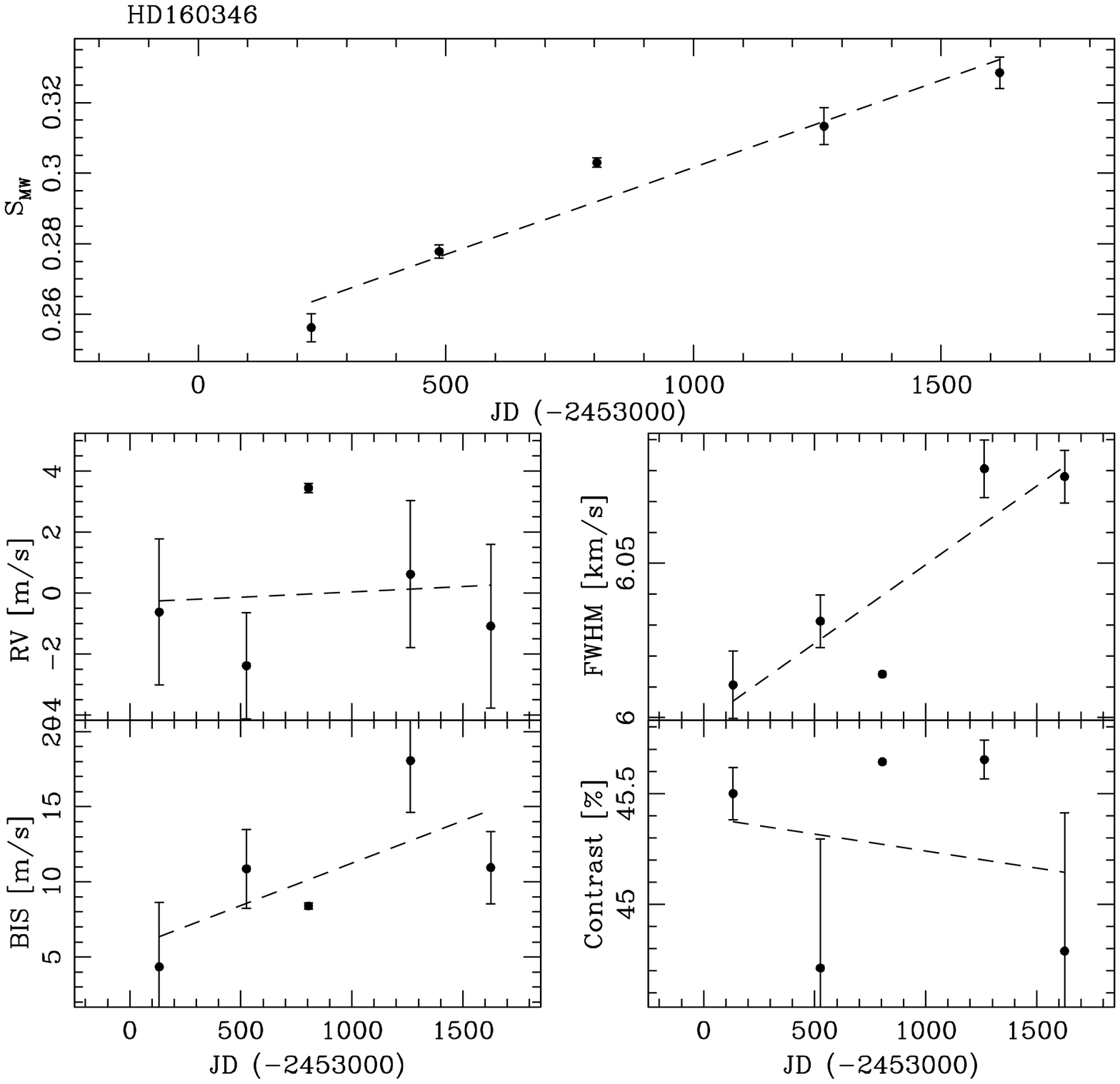}}
\resizebox{7.7truecm}{!}{\includegraphics[bb = 25 170 586 707]{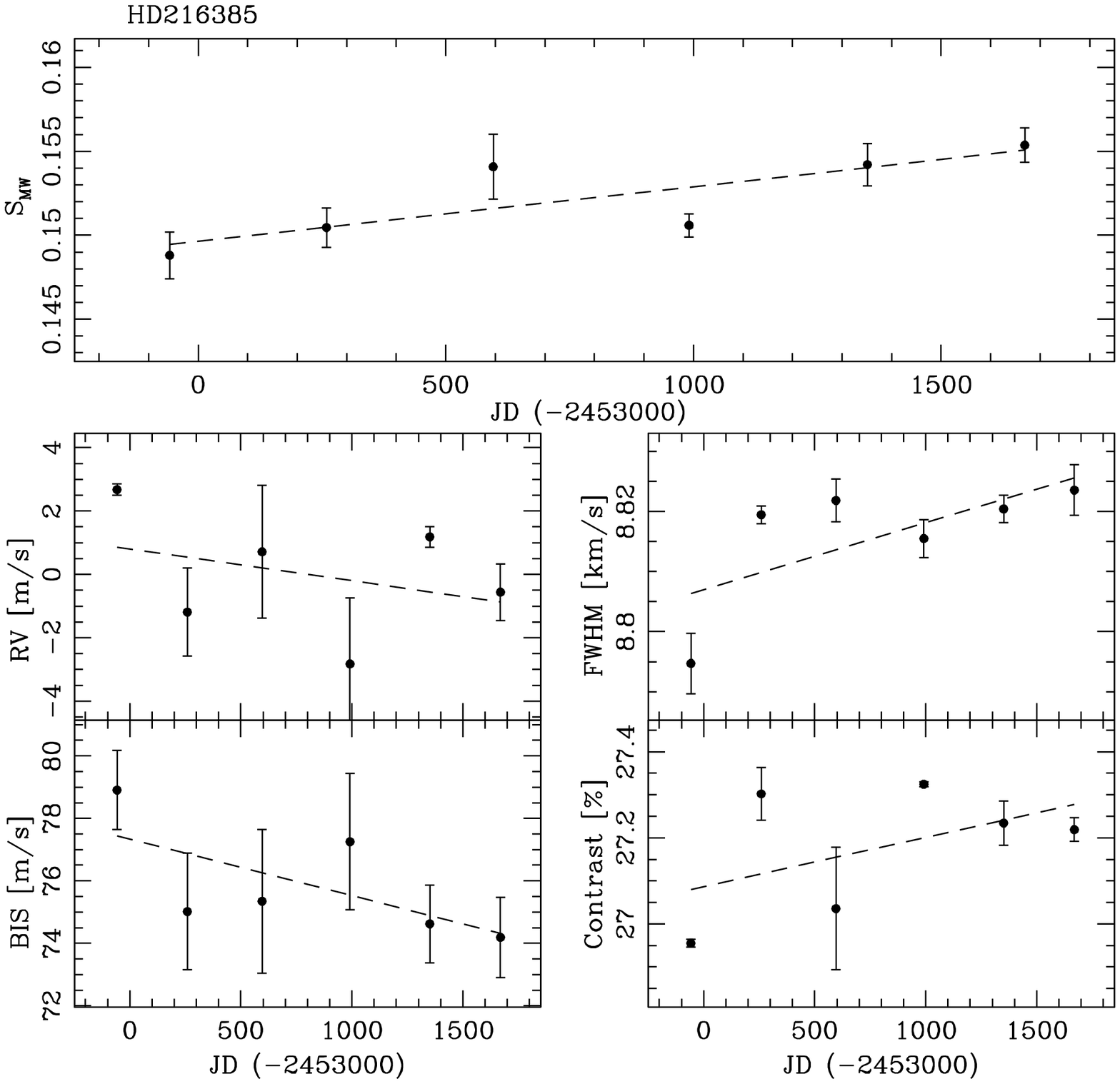}}
\caption{Yearly average S$_{MW}$ (in each top panel), radial-velocity, BIS, and the Cross-Correlation Function's Contrast and FWHM function of time for HD\,4628, HD\,16160\, and HD26965A, HD\,321467, HD\,160346, and HD\,HD216385. For comparison reasons the x-scale was set constant in all the plots. } 
\label{fig:comp1}
\end{figure*}

\subsubsection{Activity and CCF parameters}

In Fig.\,\ref{fig:comp1}  we present the yearly average time series of the S$_{MW}$, radial-velocity, BIS, and the CCF's contrast and FWHM\footnote{Data available in online Table\,8}. In the plot, the dashed lines represent linear fits to the data. Each point in the plot corresponds to the average of the 
values in time bins corresponding to the years 2003 to 2008. The error bars are computed as the rms/$\sqrt{N}$, where N is the number of
measurements in each window. The average Julian Date for each bin was considered for plotting reasons. The third date for HD\,160346 
(corresponding to the year 2006) has no error bar for FWHM and contrast, since only one measurement of this star was available in that year 
(no error bars are provided by the HARPS pipeline for these two parameters). For HD\,32147, none of the available spectra obtained in 2003 was good enough for deriving a reliable ``S'' value, though radial-velocity measurements exist.

The first remarkable result that comes out of the plots is that in most cases there seems to be a clear
positive correlation between the S$_{MW}$ index and both the FWHM (the only exceptions are HD\,32147 and our ``standard'' HD\,216385) and the BIS
of the HARPS CCF. In other words,
the FWHM and the BIS seem to be good indicators of the stellar activity level. The opposite trend (anti-correlation) is seen 
between S$_{MW}$ and the CCF's contrast. For HD\,26965\,A these relations are particularly clear since the curvature observed
in the S$_{MW}$ variation is also clearly observed in the different CCF parameters.

The reason for these facts is not {fully} clear, but they suggest that these CCF parameters can be used
as a proxy for the chromospheric activity level of a solar-type star. In particular, they can be
used to follow the activity cycle variations of the targets in complement to measurements of
different activity indexes.

\citet[][]{Boisse-2009} have shown that the parameters of the CCF, in particular its contrast, vary
as a function of the activity level of the star for the very active planet host \object{HD\,189733}. The CCF appears to be shallower 
when the star is more active. These authors explain this variation by
dark spots (typically associated with stellar activity) that have a spectrum that is different from the 
one emmited by the remaining stellar photosphere. These features
will thus naturally induce changes in the overall spectrum that are noticeable in the shape of
the CCF. {
We should note, however, that spots also influence the wings of spectral lines (decreasing their intensity),
thus decreasing their FWHM. In this case an anti-correlation between FWHM and activity level could
be expected.
}

The results presented here confirm the conclusions of \citet[][]{Boisse-2009}. Interestingly,
however, our results further suggest that these changes are detectable even in inactive (solar-like) stars,
such as the 6 objects discussed in this section.

{
Variations in the CCF contrast and FWHM can also be induced when the cores of stronger lines start 
to fill as the star becomes more active. This will make the lines (and consequently the CCF) shallower, and since
the FWHM is measured at a lower level (closer to continuum), its value will also increase.
}

We add that a change in the shape of the CCF can also be induced by 
variations in the convection pattern \citep[][]{Dravins-1982}. Observations show that solar line bisectors smaller lower velocity 
spans and convective blueshifts in active regions \citep[][]{Livingston-1982,Brandt-1990}.
This could explain the observed variations in the shape of the CCF as a function of the activity level of the star.

\subsubsection{Activity and radial velocity}

If on the one side activity cycle variations seem to induce clear signals on the above discussed  CCF parameters (FWHM, Contrast, and BIS), the 
situation regarding the RV is not so clear. For HD\,26965\,A we find a positive correlation between the ``S$_{MW}$" index (that varied by $\sim$0.04)
and the RV. Interestingly, there is a slight indication that a change in slope of the RV time series exists accompanying the observed variation in ``S''. 
The same positive correlation is observed for HD\,4628, though in this case the RV seems to present a higher dispersion. This may
also be caused by HD\,4628 presenting a lower amplitude variation in its activity level ($\Delta\,S\sim$0.02, between the first and
last years of measurements).

The opposite trend, however, is found for HD\,16160, where the ``S" index increased monotonically by about 0.054 during the period of our measurements. For this case we cannot fully exclude that a slightly wrong orbital fit  could have left any trend in the residuals. 

For HD\,32147 ($\Delta\,S\sim$0.04) and HD\,160346 ($\Delta\,S\sim$0.07), both showing clear and monotonic variation in the ``S'' index,  no significant RV variation is observed in our data. Finally, for HD\,216385, only a small ``S$_{MW}$" variation is observed (0.006). We note that this star shows very small long term chromospheric activity variation. It was actually included in our sample for its constancy in S$_{MW}$ \citep[][]{Baliunas-1995}. This is also the only F-type star in our sample.

As a test, we computed the Spearman rank correlation coefficient for the ``S$_{MW}$" vs. RV relation. A Monte-Carlo test, where we
generated random RV and ``S$_{MW}$" values with the same rms as the original sample shows that, in most cases,
a higher correlation can be found in more than 10\% of the random samplings. The only exception is HD\,26965\,A, for which
a false alarm probability of 1.5\% exists. The fact that we only have a few data points precludes any firm conclusions on this.

In face of the present (and contradictory) results, we cannot conclude anything about the existence of radial-velocity variations 
induced by long-term changes in the
chromospheric activity level in our sample of {early-K dwarfs}. More data is clearly needed, and in particular we need to cover the whole magnetic cycles
of the target stars better. The study of a larger sample could also be important. In any case, our results suggest that long-term variations 
significantly above $\sim$1\,m\,s$^{-1}$ can be excluded in
our targets as being caused by variations in the chromospheric activity level during the stellar magnetic cycle. 

We also cannot exclude the possibility that some of the observed RV variations are being caused by the presence of long-period companions,
either planets or stellar in nature. A deep adaptive optics search for close companions to our targets may be crucial for
clarifying this aspect.

\section{Discussion and conclusions}
\label{sec:conclusions}

In this paper we present the results of a long-term project to investigate the effect that long-term
stellar magnetic cycles may have on the measurement of precise radial velocities. 
For this we observed a sample of 7 late-G or early-K dwarfs and one late F dwarf (an activity ``standard'')
for more than 5 years with the HARPS spectrograph. The obtained spectra allowed us to derive a
precise value from each spectrum for the RV, a measurement of the chomospheric activity level using 3 different
spectral indexes (CaII ``S'', H$_\alpha$, and \ion{He}{i}), as well as several parameters of the CCF (BIS, FWHM, and contrast). The results of this survey suggest that in our sample:

\begin{itemize}

\item All the three activity indexes measured are good tracers of stellar activity level variations along the magnetic
cycle, though the CaII ``S'' index presents the smallest dispersion. In general, the CaII ``S'' and H$_\alpha$ indexes are correlated
with each other, while the \ion{He}{i} is anti-correlated with the first two. 

\item The activity level variations are also clearly reflected in the values of BIS, FWHM, and contrast of the HARPS CCF.
The values of BIS are particularly sensitive to activity level variations. This suggests that measurements of these parameters
can be used to clearly diagnose long-term variations in the chromospheric level of a star. The measurement of BIS, FWHM, and contrast
may even likely be useful to correct the velocity measurements for the effect of long-term stellar activity \citep[][]{Saar-2000},
in a similar way to one used by \citet[][]{Melo-2007} to correct for higher frequency noise. 
These results may be of extreme importance for present and future high-precision RV planet searches.

\item Although some of our targets show hints of low-amplitude (at the $\sim$1\,m\,s$^{-1}$ level) and long-term RV variations that could be
caused by variations in the activity level of the star, our results are not conclusive about the nature and amplitude of this effect. 
Our data suggests that for {early-K} dwarfs the variation in the stellar activity level along the magnetic cycles does 
not strongly induce variations in the measured RVs. {Early-K} dwarfs are thus {\it bona fide} targets to search for
very low-mass planets using precise radial velocity instruments. We note that these targets already have the lowest
granulation and oscillation ``noise'' level among solar type stars (Dumusque et al., in preparation). 
\end{itemize}

A better understanding of the influence of long-term stellar magnetic cycles on the radial-velocities is certainly crucial to
plan future radial-velocity planet searches with a new generation of instruments like ESPRESSO@VLT\footnote{http://espresso.astro.up.pt}
or CODEX@E-ELT \citep[][]{Pasquini-2008}. These instruments will be able to obtain RVs with a precision of better than 10\,cm\,s$^{-1}$,
allowing the detection of Earth-like planets in the habitable zone of nearly solar-type stars. 

To further investigate the above discussed issues, a continuation of the present program is needed. Furthermore, it would be very important
to extend our sample to earlier type G-dwarfs (more similar to our Sun). Convective blueshifts are lower in K-dwarfs than in
F- and G-dwarfs \citep[][]{Gray-1992}. In active regions  the convective velocities are lower, also implying smaller line asymmetries \citep[][]{Livingston-1982,Brandt-1990}. We may thus expect a stronger effect of chromospheric activity for earlier type dwarfs {\citep[e.g.][]{Saar-2009}. }

As a complement, it would also be important to investigate the existence of long period stellar companions in detail, which could
be able to induce long-term RV variations. An adaptive optics survey should thus be conducted to investigate this possibility.

\begin{acknowledgements}
We would like to thank our referee, S. Saar, for the very positive and helpful report. NCS would like to acknowledge the support by the European Research Council/European Community under the FP7 through a starting grant, as well from Funda\c{c}\~ao para a Ci\^encia e a Tecnologia (FCT), Portugal, through program Ci\^encia\,2007, and in the form of grants PTDC/CTE-AST/098528/2008 and PTDC/CTE-AST/098604/2008. 
JGS would like to acknowledge the support by EC's FP6 and by FCT (with POCI2010 and FEDER funds), within the HELAS international collaboration.
\end{acknowledgements}

\bibliographystyle{aa}
\bibliography{santos_bibliography}

\begin{thebibliography}{52}
\expandafter\ifx\csname natexlab\endcsname\relax\def\natexlab#1{#1}\fi

\bibitem[{{Baliunas} {et~al.}(1996){Baliunas}, {Sokoloff}, \&
  {Soon}}]{Baliunas-1996}
{Baliunas}, S., {Sokoloff}, D., \& {Soon}, W. 1996, ApJ, 457, L99+

\bibitem[{{Baliunas} {et~al.}(1995){Baliunas}, {Donahue}, {Soon}, {Horne},
  {Frazer}, {Woodard-Eklund}, {Bradford}, {Rao}, {Wilson}, {Zhang}, {Bennett},
  {Briggs}, {Carroll}, {Duncan}, {Figueroa}, {Lanning}, {Misch}, {Mueller},
  {Noyes}, {Poppe}, {Porter}, {Robinson}, {Russell}, {Shelton}, {Soyumer},
  {Vaughan}, \& {Whitney}}]{Baliunas-1995}
{Baliunas}, S.~L., {Donahue}, R.~A., {Soon}, W.~H., {et~al.} 1995, ApJ, 438,
  269

\bibitem[{{Boisse} {et~al.}(2009){Boisse}, {Moutou}, {Vidal-Madjar}, {Bouchy},
  {Pont}, {H{\'e}brard}, {Bonfils}, {Croll}, {Delfosse}, {Desort}, {Forveille},
  {Lagrange}, {Loeillet}, {Lovis}, {Matthews}, {Mayor}, {Pepe}, {Perrier},
  {Queloz}, {Rowe}, {Santos}, {S{\'e}gransan}, \& {Udry}}]{Boisse-2009}
{Boisse}, I., {Moutou}, C., {Vidal-Madjar}, A., {et~al.} 2009, \aap, 495, 959

\bibitem[{{Bonfils} {et~al.}(2007){Bonfils}, {Mayor}, {Delfosse}, {Forveille},
  {Gillon}, {Perrier}, {Udry}, {Bouchy}, {Lovis}, {Pepe}, {Queloz}, {Santos},
  \& {Bertaux}}]{Bonfils-2007}
{Bonfils}, X., {Mayor}, M., {Delfosse}, X., {et~al.} 2007, ArXiv e-prints, 704

\bibitem[{{Bouchy} {et~al.}(2004){Bouchy}, {Pont}, {Santos}, {Melo}, {Mayor},
  {Queloz}, \& {Udry}}]{Bouchy-2004}
{Bouchy}, F., {Pont}, F., {Santos}, N.~C., {et~al.} 2004, A\&A, 421, L13

\bibitem[{{Brandt} \& {Solanki}(1990)}]{Brandt-1990}
{Brandt}, P.~N. \& {Solanki}, S.~K. 1990, A\&A, 231, 221

\bibitem[{{Cincunegui} {et~al.}(2007){Cincunegui}, {D{\'{\i}}az}, \&
  {Mauas}}]{Cincunegui-2007}
{Cincunegui}, C., {D{\'{\i}}az}, R.~F., \& {Mauas}, P.~J.~D. 2007, \aap, 469,
  309

\bibitem[{{Danks} \& {Lambert}(1985)}]{Danks-1985}
{Danks}, A.~C. \& {Lambert}, D.~L. 1985, A\&A, 148, 293

\bibitem[{{Dravins}(1982)}]{Dravins-1982}
{Dravins}, D. 1982, ARA\&A, 20, 61

\bibitem[{{Duquennoy} \& {Mayor}(1991)}]{Duquennoy-1991}
{Duquennoy}, A. \& {Mayor}, M. 1991, A\&A, 248, 485

\bibitem[{{ESA}(1997)}]{ESA-1997}
{ESA}. 1997, The Hipparcos and Tycho Catalogues

\bibitem[{{Fernandes} \& {Santos}(2004)}]{Fernandes-2004}
{Fernandes}, J. \& {Santos}, N.~C. 2004, A\&A, 427, 607

\bibitem[{{Flower}(1996)}]{Flower-1996}
{Flower}, P.~J. 1996, ApJ, 469, 355

\bibitem[{{Golimowski} {et~al.}(2000){Golimowski}, {Henry}, {Krist},
  {Schroeder}, {Marcy}, {Fischer}, \& {Butler}}]{Golimowski-2000}
{Golimowski}, D.~A., {Henry}, T.~J., {Krist}, J.~E., {et~al.} 2000, AJ, 120,
  2082

\bibitem[{{Gray}(1992)}]{Gray-1992}
{Gray}, D.~F. 1992, {The Observation and Analysis of Stellar Photospheres} (The
  Observation and Analysis of Stellar Photospheres, by David F.~Gray,
  pp.~470.~ISBN 0521408687.~Cambridge, UK: Cambridge University Press, June
  1992.)

\bibitem[{{Halbwachs} {et~al.}(2003){Halbwachs}, {Mayor}, {Udry}, \&
  {Arenou}}]{Halbwachs-2003}
{Halbwachs}, J.~L., {Mayor}, M., {Udry}, S., \& {Arenou}, F. 2003, A\&A, 397,
  159

\bibitem[{{Heintz}(1974)}]{Heintz-1974}
{Heintz}, W.~D. 1974, AJ, 79, 819

\bibitem[{{Hu{\'e}lamo} {et~al.}(2008){Hu{\'e}lamo}, {Figueira}, {Bonfils},
  {Santos}, {Pepe}, {Gillon}, {Azevedo}, {Barman}, {Fern{\'a}ndez}, {di Folco},
  {Guenther}, {Lovis}, {Melo}, {Queloz}, \& {Udry}}]{Huelamo-2008}
{Hu{\'e}lamo}, N., {Figueira}, P., {Bonfils}, X., {et~al.} 2008, A\&A, 489, L9

\bibitem[{{K{\"u}rster} {et~al.}(2003){K{\"u}rster}, {Endl}, {Rouesnel}, {Els},
  {Kaufer}, {Brillant}, {Hatzes}, {Saar}, \& {Cochran}}]{Kurster-2003}
{K{\"u}rster}, M., {Endl}, M., {Rouesnel}, F., {et~al.} 2003, A\&A, 403, 1077

\bibitem[{{Kurucz}(1993)}]{Kurucz-1993}
{Kurucz}, R. 1993, ATLAS9 Stellar Atmosphere Programs and 2 km/s grid.~Kurucz
  CD-ROM No.~13.~ Cambridge, Mass.: Smithsonian Astrophysical Observatory,
  1993., 13

\bibitem[{{Landman}(1981)}]{Landman-1981}
{Landman}, D.~A. 1981, ApJ, 244, 345

\bibitem[{{L{\'e}pine} \& {Bongiorno}(2007)}]{Lepine-2007}
{L{\'e}pine}, S. \& {Bongiorno}, B. 2007, AJ, 133, 889

\bibitem[{{Livingston} {et~al.}(2007){Livingston}, {Wallace}, {White}, \&
  {Giampapa}}]{Livingston-2007}
{Livingston}, W., {Wallace}, L., {White}, O.~R., \& {Giampapa}, M.~S. 2007,
  ApJ, 657, 1137

\bibitem[{{Livingston}(1982)}]{Livingston-1982}
{Livingston}, W.~C. 1982, Nature, 297, 208

\bibitem[{{Mayor} {et~al.}(2009){Mayor}, {Bonfils}, {Forveille}, {Delfosse},
  {Udry}, {Bertaux}, {Beust}, {Bouchy}, {Lovis}, {Pepe}, {Perrier}, {Queloz},
  \& {Santos}}]{Mayor-2009}
{Mayor}, M., {Bonfils}, X., {Forveille}, T., {et~al.} 2009, ArXiv e-prints

\bibitem[{{Mayor} \& {Queloz}(1995)}]{Mayor-1995}
{Mayor}, M. \& {Queloz}, D. 1995, Nature, 378, 355

\bibitem[{{McMillan} {et~al.}(1993){McMillan}, {Moore}, {Perry}, \&
  {Smith}}]{McMillan-1993}
{McMillan}, R.~S., {Moore}, T.~L., {Perry}, M.~L., \& {Smith}, P.~H. 1993, ApJ,
  403, 801

\bibitem[{{Melo} {et~al.}(2007){Melo}, {Santos}, {Gieren}, {Pietrzynski},
  {Ruiz}, {Sousa}, {Bouchy}, {Lovis}, {Mayor}, {Pepe}, {Queloz}, {Da Silva}, \&
  {Udry}}]{Melo-2007}
{Melo}, C., {Santos}, N.~C., {Gieren}, W., {et~al.} 2007, A\&A, 467, 721

\bibitem[{{Meunier} \& {Delfosse}(2009)}]{Meunier-2009}
{Meunier}, N. \& {Delfosse}, X. 2009, A\&A, 501, 1103

\bibitem[{{Noyes} {et~al.}(1984){Noyes}, {Hartmann}, {Baliunas}, {Duncan}, \&
  {Vaughan}}]{Noyes-1984}
{Noyes}, R.~W., {Hartmann}, L.~W., {Baliunas}, S.~L., {Duncan}, D.~K., \&
  {Vaughan}, A.~H. 1984, ApJ, 279, 763

\bibitem[{{Pasquini} {et~al.}(2008){Pasquini}, {Avila}, {Delabre}, {Dekker},
  {D'Odorico}, {Liske}, {Manescau}, {Bonifacio}, {Cristiani}, {D'Odorico},
  {Molaro}, {Vanzella}, {Santin}, {Viel}, {Dessauges-Zavadsky}, {Lovis},
  {Mayor}, {Pepe}, {Queloz}, {Udry}, {Haehnelt}, {Murphy}, {Garcia-Lopez},
  {Bouchy}, {Levshakov}, \& {Zucker}}]{Pasquini-2008}
{Pasquini}, L., {Avila}, G., {Delabre}, B., {et~al.} 2008, in Precision
  Spectroscopy in Astrophysics, ed. N.~C. {Santos}, L.~{Pasquini}, A.~C.~M.
  {Correia}, \& M.~{Romaniello}, 249--253

\bibitem[{{Paulson} {et~al.}(2002){Paulson}, {Saar}, {Cochran}, \&
  {Hatzes}}]{Paulson-2002}
{Paulson}, D.~B., {Saar}, S.~H., {Cochran}, W.~D., \& {Hatzes}, A.~P. 2002, AJ,
  124, 572

\bibitem[{{Pepe} {et~al.}(2002){Pepe}, {Mayor}, {Rupprecht}, {Avila},
  {Ballester}, {Beckers}, {Benz}, {Bertaux}, {Bouchy}, {Buzzoni}, {Cavadore},
  {Deiries}, {Dekker}, {Delabre}, {D'Odorico}, {Eckert}, {Fischer}, {Fleury},
  {George}, {Gilliotte}, {Gojak}, {Guzman}, {Koch}, {Kohler}, {Kotzlowski},
  {Lacroix}, {Le Merrer}, {Lizon}, {Lo Curto}, {Longinotti}, {Megevand},
  {Pasquini}, {Petitpas}, {Pichard}, {Queloz}, {Reyes}, {Richaud}, {Sivan},
  {Sosnowska}, {Soto}, {Udry}, {Ureta}, {van Kesteren}, {Weber}, {Weilenmann},
  {Wicenec}, {Wieland}, {Christensen-Dalsgaard}, {Dravins}, {Hatzes}, {K{\"
  u}rster}, {Paresce}, \& {Penny}}]{Pepe-2002}
{Pepe}, F., {Mayor}, M., {Rupprecht}, G., {et~al.} 2002, The Messenger, 110, 9

\bibitem[{{Queloz} {et~al.}(2001){Queloz}, {Henry}, {Sivan}, {Baliunas},
  {Beuzit}, {Donahue}, {Mayor}, {Naef}, {Perrier}, \& {Udry}}]{Queloz-2001}
{Queloz}, D., {Henry}, G.~W., {Sivan}, J.~P., {et~al.} 2001, A\&A, 379, 279

\bibitem[{{Queloz} {et~al.}(2000){Queloz}, {Mayor}, {Weber}, {Bl{\' e}cha},
  {Burnet}, {Confino}, {Naef}, {Pepe}, {Santos}, \& {Udry}}]{Queloz-2000}
{Queloz}, D., {Mayor}, M., {Weber}, L., {et~al.} 2000, A\&A, 354, 99

\bibitem[{{Saar}(2009)}]{Saar-2009}
{Saar}, S.~H. 2009, in American Institute of Physics Conference Series, Vol.
  1094, American Institute of Physics Conference Series, ed. {E.~Stempels},
  152--161

\bibitem[{{Saar} \& {Donahue}(1997)}]{Saar-1997}
{Saar}, S.~H. \& {Donahue}, R.~A. 1997, ApJ, 485, 319

\bibitem[{{Saar} \& {Fischer}(2000)}]{Saar-2000}
{Saar}, S.~H. \& {Fischer}, D. 2000, ApJ, 534, L105

\bibitem[{{Saar} {et~al.}(1997){Saar}, {Huovelin}, {Osten}, \&
  {Shcherbakov}}]{Saar-1997b}
{Saar}, S.~H., {Huovelin}, J., {Osten}, R.~A., \& {Shcherbakov}, A.~G. 1997,
  A\&A, 326, 741

\bibitem[{{Santos} {et~al.}(2004{\natexlab{a}}){Santos}, {Bouchy}, {Mayor},
  {Pepe}, {Queloz}, {Udry}, {Lovis}, {Bazot}, {Benz}, {Bertaux}, {Lo Curto},
  {Delfosse}, {Mordasini}, {Naef}, {Sivan}, \& {Vauclair}}]{Santos-2004a}
{Santos}, N.~C., {Bouchy}, F., {Mayor}, M., {et~al.} 2004{\natexlab{a}}, A\&A,
  426, L19

\bibitem[{{Santos} {et~al.}(2004{\natexlab{b}}){Santos}, {Israelian}, \&
  {Mayor}}]{Santos-2004b}
{Santos}, N.~C., {Israelian}, G., \& {Mayor}, M. 2004{\natexlab{b}}, A\&A, 415,
  1153

\bibitem[{{Santos} {et~al.}(2000){Santos}, {Mayor}, {Naef}, {Pepe}, {Queloz},
  {Udry}, \& {Blecha}}]{Santos-2000a}
{Santos}, N.~C., {Mayor}, M., {Naef}, D., {et~al.} 2000, A\&A, 361, 265

\bibitem[{{Schaerer} {et~al.}(1993{\natexlab{a}}){Schaerer}, {Charbonnel},
  {Meynet}, {Maeder}, \& {Schaller}}]{Schaerer-1993a}
{Schaerer}, D., {Charbonnel}, C., {Meynet}, G., {Maeder}, A., \& {Schaller}, G.
  1993{\natexlab{a}}, A\&AS, 102, 339

\bibitem[{{Schaerer} {et~al.}(1993{\natexlab{b}}){Schaerer}, {Meynet},
  {Maeder}, \& {Schaller}}]{Schaerer-1993b}
{Schaerer}, D., {Meynet}, G., {Maeder}, A., \& {Schaller}, G.
  1993{\natexlab{b}}, A\&AS, 98, 523

\bibitem[{{Schaller} {et~al.}(1992){Schaller}, {Schaerer}, {Meynet}, \&
  {Maeder}}]{Schaller-1992}
{Schaller}, G., {Schaerer}, D., {Meynet}, G., \& {Maeder}, A. 1992, A\&AS, 96,
  269

\bibitem[{{Sneden}(1973)}]{Sneden-1973}
{Sneden}, C. 1973, Ph.D. Thesis, Univ. of Texas

\bibitem[{{Sousa} {et~al.}(2008){Sousa}, {Santos}, {Mayor}, {Udry},
  {Casagrande}, {Israelian}, {Pepe}, {Queloz}, \& {Monteiro}}]{Sousa-2008}
{Sousa}, S.~G., {Santos}, N.~C., {Mayor}, M., {et~al.} 2008, A\&A, 487, 373

\bibitem[{{Tokovinin}(1997)}]{Tokovinin-1997}
{Tokovinin}, A.~A. 1997, A\&AS, 124, 75

\bibitem[{{Udry} \& {Santos}(2007)}]{Udry-2007}
{Udry}, S. \& {Santos}, N. 2007, ARAA, 45, 397

\bibitem[{{Vaughan} \& {Preston}(1980)}]{Vaughan-1980}
{Vaughan}, A.~H. \& {Preston}, G.~W. 1980, PASP, 92, 385

\bibitem[{{Vaughan} {et~al.}(1978){Vaughan}, {Preston}, \&
  {Wilson}}]{Vaughan-1978}
{Vaughan}, A.~H., {Preston}, G.~W., \& {Wilson}, O.~C. 1978, PASP, 90, 267

\bibitem[{{Wright} {et~al.}(2008){Wright}, {Marcy}, {Butler}, {Vogt}, {Henry},
  {Isaacson}, \& {Howard}}]{Wright-2008}
{Wright}, J.~T., {Marcy}, G.~W., {Butler}, R.~P., {et~al.} 2008, ApJ, 683, L63

\end{thebibliography}

\end{document}